\documentclass{emulateapj}

\journalinfo{To appear in The Astrophysical Journal 20 November 2007}
\submitted{Submitted: 22 June 2007; accepted: 5 August 2007} 

\newcommand{\chandra}{{\em{Chandra}}}
\newcommand{\rosat}{{\em{ROSAT}}}
\newcommand{\cxo}{{\it{Chandra X-ray Observatory}}}
\newcommand{\xmm}{{\it{XMM-Newton}}}

\shorttitle{X-ray Flux Limits to Four Type Ia SNe}
\shortauthors{Hughes et al.}

\begin{document}

\title{\chandra\ Observations of Type Ia Supernovae: Upper Limits to the
X-ray Flux of SN~2002bo, SN~2002ic, SN~2005gj, and SN~2005ke}

\author{ John P. Hughes,\altaffilmark{1,2}
 Nikolai Chugai,\altaffilmark{3}
 Roger Chevalier,\altaffilmark{4}
 Peter Lundqvist,\altaffilmark{5}
and
 Eric Schlegel\altaffilmark{6}
}

\altaffiltext{1}{Department of Physics \& Astronomy, Rutgers University, 
136 Frelinghuysen Road, Piscataway, NJ 08854-8019; jph@physics.rutgers.edu}
\altaffiltext{2}{Also Department of Astrophysical Sciences, Princeton 
University,  Peyton Hall,  Princeton, NJ 08544-1001}
\altaffiltext{3}{Institute of Astronomy, RAS, Pyatnitskaya 48, 109017 
Moscow, Russia; nchugai@inasan.ru}
\altaffiltext{4}{Department of Astronomy, University of Virginia, 
P.O.~Box 400325, Charlottesville, VA 22904; rac5x@astsun.astro.virginia.edu}
\altaffiltext{5}{Stockholm Observatory, AlbaNova, Department of 
Astronomy, SE-106 91 Stockholm, Sweden; peter@astro.su.se}
\altaffiltext{6}{Department of Physics \& Astronomy, The University of 
Texas at San Antonio, One UTSA Circle, San Antonio, Texas 78249;
eric.schlegel@utsa.edu}

\begin{abstract}

We set sensitive upper limits to the X-ray emission of four Type Ia
supernovae (SNe~Ia) using the \cxo.  SN~2002bo, a normal, although
reddened, nearby SN~Ia, was observed 9.3 days after explosion. For an
absorbed, high temperature bremsstrahlung model the flux limits are
$3.2\times 10^{-16}$ ergs cm$^{-2}$ s$^{-1}$ (0.5--2 keV band) and
$4.1\times 10^{-15}$ ergs cm$^{-2}$ s$^{-1}$ (2--10 keV band). Using
conservative model assumptions and a 10 km s$^{-1}$ wind speed, we
derive a mass loss rate of $\dot{M} \sim 2\times 10^{-5}\, M_\odot$
yr$^{-1}$, which is comparable to limits set by the non-detection of
H$\alpha$ lines from other SNe~Ia.  Two other objects, SN~2002ic and
SN~2005gj, observed 260 and 80 days after explosion, respectively, are
the only SNe~Ia showing evidence for circumstellar interaction.  The
SN~2002ic X-ray flux upper limits are $\sim$4 times below predictions
of the interaction model currently favored to explain the bright
optical emission.  To resolve this discrepancy we invoke the mixing of
cool dense ejecta fragments into the forward shock region, which
produces increased X-ray absorption. A modest amount of mixing allows
us to accommodate the \chandra\ upper limit. 
SN~2005gj is less well studied at this time. Assuming the same
circumstellar environment as for SN~2002i, the X-ray flux upper limits
for SN~2005gj are $\sim$4 times below the predictions, suggesting
that mixing of cool ejecta into the forward shock has also occurred here.
Our reanalysis of Swift and \chandra\ data on SN~2005ke does not confirm
a previously reported X-ray detection.  The host galaxies NGC~3190
(SN~2002bo) and NGC~1371 (SN~2005ke) each harbor a low luminosity
($L_X \sim 3-4 \times 10^{40}$ ergs s$^{-1}$) active nucleus in
addition to wide-spread diffuse soft X-ray emission.

\end{abstract}


\section{Introduction}

Type Ia supernovae (SN~Ia) are an important subclass of supernova (SN)
that are thought to arise from explosions of white dwarfs in binary
systems, although the exact nature of their progenitor systems is
largely a mystery.  This ignorance is not due to a lack of effort
since SN~Ia are the subject of intense scrutiny, not least of all
because of their importance as cosmological probes. SN~Ia have been
used to measure the Hubble constant \citep{ham95, riess96} and have
even given strong evidence for a non-zero cosmological constant
\citep{riess98, perl99}.  Our failure to identify SN~Ia progenitors
highlights a major gap in our understanding of stellar evolution in
binary systems, and presents a stumbling-block to understanding the
chemical evolution of galaxies, since Ia's are known to be efficient
producers of iron.  In addition, until SN~Ia progenitor systems are
clearly identified, it will remain difficult to convincingly eliminate
evolutionary sources of systematic error in the observed trend of SN brightness
with redshift.

The current model framework for Type Ia SNe involves carbon
deflagration/detonation in a white dwarf driven close to the Chandrasekhar
limit by accretion. The most likely type of progenitor system
\citep{branch95} is a C-O white dwarf accreting H/He-rich gas from a
companion, either from its wind or through Roche lobe overflow.
Double degenerate scenarios with coalescing pairs of C-O white dwarfs
are also possible, while sub-Chandrasekhar-mass white dwarfs are less
likely \citep{branch95}.  In the non-coalescing scenarios, there will
be circumstellar gas whose composition and geometry depend on the
nature of the progenitor system.  If the circumstellar medium (CSM)
emits radiation, or absorbs radiation from the SN, this can be used to
distinguish between types of progenitor systems.

Thermal X-ray emission is expected to arise from the hot gas in the
interaction region between the rapidly moving supernova ejecta and the
wind from presupernova mass loss.  Detecting this X-ray emission
offers a direct and potentially quite sensitive probe of the amount of
CSM.  Because of its high sensitivity, broad bandwidth, unprecedented
imaging capability, and ability to respond rapidly to a target of
opportunity (ToO) request, the \cxo\ is the premier instrument to
search for faint X-ray emission from transient point sources.

In the following we present the analysis and interpretation of
\chandra\ observations of four recent SNe~Ia. Two are fairly normal
SNe~Ia, while the others are peculiar cases showing strong H$\alpha$
emission from circumstellar interaction.  The following section
describes the observed targets, \S3 presents the observations and
techniques, \S4 uses the upper limits on X-ray flux to constrain the
nature of the CSM, and the final section concludes.  In an Appendix we
describe the X-ray properties of the host galaxies for the two nearby
systems, NGC~3190 and NGC~1371.

\section{Description of Targets}

Our first targeted SN, SN~2002bo, was discovered on 2002 March 9
\citep{cace02} and was spectroscopically confirmed as a Type Ia SN
shortly thereafter \citep{kawa02,bene02,math02}.  The SN appeared
approximately 18$^{\prime\prime}$ southeast of the nucleus of NGC~3190
in the middle of an apparent dust lane. Early optical spectra of the
SN indicated that it was quite young (10--14 days before maximum) and
possibly reddened.  Based on the SN's proximity to Earth and very
early discovery, we triggered a pre-approved \chandra\ ToO observation
on March 12.  The nominally 20 ks long observation was carried out two
days later (the midpoint of the observation was at JD 2452347.9).  We
now know that the time of maximum light (in the B band) occurred on
2002 March 23.0 UT or, equivalently JD 2452356.5, \citep{bene04,
kris04} and that the inferred explosion epoch was JD 2452338.6 $\pm$
0.5.  Our \chandra\ ToO observation was therefore made only 9.3 days
after explosion.  We adopt a value of 22 Mpc for the distance to
NGC~3190 \citep[for more details, see][]{bene04, kris04}. SN~2002bo
showed significant reddening $E(B-V) = 0.43 \pm 0.10$ \citep{bene04},
most likely due to its host galaxy.  We converted this to an
equivalent X-ray absorbing column value of $3 \times 10^{21}\, \rm
cm^{-2}$ using \citet[p.~527]{cox00} in order to model the SN's X-ray
flux.

The second SN~Ia we observed with \chandra, SN~2002ic, is the first SN~Ia
in which a circumstellar interaction was detected \citep[however,
see][for an alternate view]{bene06}.  Discovered on 2002
November 13 \citep{wood02}, it was spectroscopically identified on 2002
December 7 as a type Ia SN at a redshift of 0.0666 \citep{ham02}.  In
2003 June \citet{ham03a} reported the detection of broad (FWHM $\sim$
1800 km s$^{-1}$) H$\alpha$ emission from the supernova, revealing a
SN/CSM interaction in SN~2002ic involving a dense CSM with as much as
several solar masses of material \citep{ham03b}.  Evidence has been
presented for a clumpy, aspherical structure for the CSM
\citep{wang04,deng04,chuche07} as well as for gaps or troughs in the radial
profile \citep{wood04,chuyun04}.  The unprecedented strength of the CSM
interaction in SN~2002ic combined with the persistence of the H$\alpha$
emission (which lasted for many months), led us to request a 20-ks
observation with \chandra\ under the Director's discretionary time
program.  The request was approved and our observation was carried out
on JD 2452863.5 (the observation began on 2003 August 11).
\citet{wood04} estimate a date for maximum light in the B-band of JD
2452606.  Assuming a typical time between explosion and maximum light
of $\sim$18 days \citep{riess99} we estimate an explosion date of JD
2452588.  The \chandra\ observation was made 275 days after explosion
or approximately 260 days in the SN frame.

Another peculiar SN~Ia (SN~2005gj), showing a circumstellar interaction,
was discovered on 2005 September 26 by the SDSS-II collaboration
\citep{barentine05,frieman05}. Originally classified as a probable
type-IIn supernova because of the presence of resolved Hydrogen line
emission, further spectroscopy demonstrated a remarkable similarity
with SN~2002ic, resulting in a reclassification of the event as a SN~Ia
but with clear signs of a circumstellar interaction \citep{prieto05}.
The SN lies in an anonymous galaxy at a redshift of 0.062
\citep{barentine05}.  No X-ray emission was detected in a brief
observation on 2005 November 24 using the X-ray telescope onboard the
Swift satellite \citep{immler05}.  A 50-ks long \chandra\ observation
was carried out under the Director's discretionary time (DDT) program
on JD 2453716.5 (the observation began on 2005 December 11).
\citet{aldetal06} recently presented an analysis of the optical light
curves and spectra in which they conservatively estimate that
SN~2005gj exploded sometime in the interval September 18.6--24.6.
More recently, \citet{prietoetal07} estimate the time of explosion to
be September $24.4\pm2.0$, based simply on whether or not they
detected the SN, which, conservatively, can only provide an upper
limit to the time of explosion.  Since a difference of a few days is
not significant for our purposes, we take the midpoint value of the
\citet{aldetal06} study (JD 2453635.1) as our estimate for the
explosion date.  This is some 77 days (in the SN frame) before the
\chandra\ observation.

The final SN~Ia we report on here is SN~2005ke, which was discovered on
2005 November 13 \citep{puckett05} and confirmed as a Type Ia SN a few
days later \citep{patat05}.  These later authors estimated the
spectral age by comparison to SN~1999by to be a few days before
maximum light, which suggests an approximate explosion date of Nov 2
(JD 2453676.50). Although the host galaxy, NGC~1371, is relatively
nearby \citep[we use a distance of 17 Mpc;][]{tully88} the SN was
underluminous and not discovered particularly early so it was not
initially observed by \chandra.  However, Swift devoted considerable
observing time to the SN ($\sim$250 ks) over the course of the
2005/2006 winter and reported a tentative X-ray detection at about the 3
$\sigma$ significance level with a flux of $\sim$$4\times 10^{-15}$
ergs cm$^{-2}$ s$^{-2}$ (0.3--2 keV band) \citep{immler06}. The
reported midpoint of the Swift observation is some 42 days after
explosion.  Subsequently a \chandra\ DDT observation was approved and
carried out on JD 2453786.1 (beginning on 2006 February 19) about 110
days after explosion.  No radio emission has been detected
\citep{soderberg06}.

Table~\ref{tab_intro} summaries some key observational information
about the several SNe discussed in this paper.

\begin{deluxetable*}{lcccc}
\tablewidth{0pt}
\tablecaption{Observed Supernovae}
\tablecolumns{5}
\tablehead{
  \colhead{} & 
  \colhead{SN~2002bo} & 
  \colhead{SN~2002ic} & 
  \colhead{SN~2005gj} & 
  \colhead{SN~2005ke} }
\startdata
 R.A.~(J2000)  & 10:18:06.5 & 01:30:02.6  & 03:01:12.0    & 03:35:04.4 \\
 Decl.~(J2000) & 21:49:41   & 21:53:07    & $-$00:33:13.9 & $-$24:56:38.8 \\
 Explosion date (JD)    &  2452338.6  & 2452588.0 & 2453635.1 & 2453676.5 \\ 
 \chandra\ Obs. date (JD) & 2452347.9 & 2452863.5 & 2453716.5 & 2453786.1 \\ 
 \chandra\ exposure (ks) &   19.6     & 17.3      &   49.4    &   14.8    \\
 $N_{{\rm H I}}$ (atoms cm$^{-2}$)
                         & $2.1 \times 10^{20}$ & $6.2 \times 10^{20}$ 
                         & $7.1 \times 10^{20}$ & $1.4 \times 10^{20}$ \\
 D (Mpc) / $z$           &  22   &  0.0666    & 0.062     &  17   \\
\enddata
\label{tab_intro}
\end{deluxetable*}

\section{Observations and Techniques}\label{s-obs}

Each target was observed near the nominal telescope aimpoint on the
back-side illuminated chip (S3) of the \chandra\ Advanced CCD Imaging
Spectrometer (hereafter ACIS-S). Although the background rate is
higher here than on the front-side illuminated chips of the imaging
array, the greatly enhanced low energy response of chip S3 was the
main deciding factor in favor of using it.  Observations were carried
out in timed exposure mode with events recorded in FAINT (SN~2002bo:
Observation Identifier [ObsID] 2760; SN~2005ke: ObsID 7277) or VFAINT
(SN~2002ic: ObsID 4449; SN~2005gj: ObsID 7241) data mode. The same basic
data reduction processes were applied to each observation, using
standard \chandra\ software and calibration (CIAO version 3.2.2, CALDB
3.1.0). This included applying the latest gain map, identifying and
removing hot pixel and afterglow events, using VFAINT mode information
(when available) to remove some non-X-ray (i.e., charged particle)
background events, and filtering on grade (retaining the usual values
02346) and status.  Light curves of the entire S3 chip were made to
identify times of high background.  The SN~2002bo, SN~2005gj, and
SN~2005ke data were mostly free of flares, while the SN~2002ic data
showed a modest amount of background flaring.  Excluding time
intervals more than 3 $\sigma$ from the mean resulted in final useful
exposure times of 19620 s (SN~2002bo), 17280 s (SN~2002ic), 49410 s
(SN~2005gj), and 14810 s (SN~2005ke).  The astrometry of each pointing
was checked as well. We found no unambiguous matches between X-ray and
optical point sources in the field of SN~2002bo, so it was not possible
to obtain an independent check of the absolute astrometry of the X-ray
data. In lieu of this the on-line ``Aspect Calculator'' tool was run,
yielding an aspect offset of 0.5$^{\prime\prime}$ which was applied to
the data.  For SN~2002ic, it was possible to match four X-ray sources
with optical counterparts in the USNO-A2.0 catalog from which an
aspect offset of 0.6$^{\prime\prime}$ was determined for the \chandra\
data.  There were nine optical counterparts to X-ray sources in the
SN~2005gj field which resulted in an aspect correction of
1.0$^{\prime\prime}$ to the \chandra\ data.  One X-ray source showed a
clear optical counterpart for the SN~2005ke observation.  The
positional offset was less than 0.1$^{\prime\prime}$ so no adjustment
was applied to the \chandra\ astrometry.

\begin{figure*}
\epsscale{.45}
\plotone{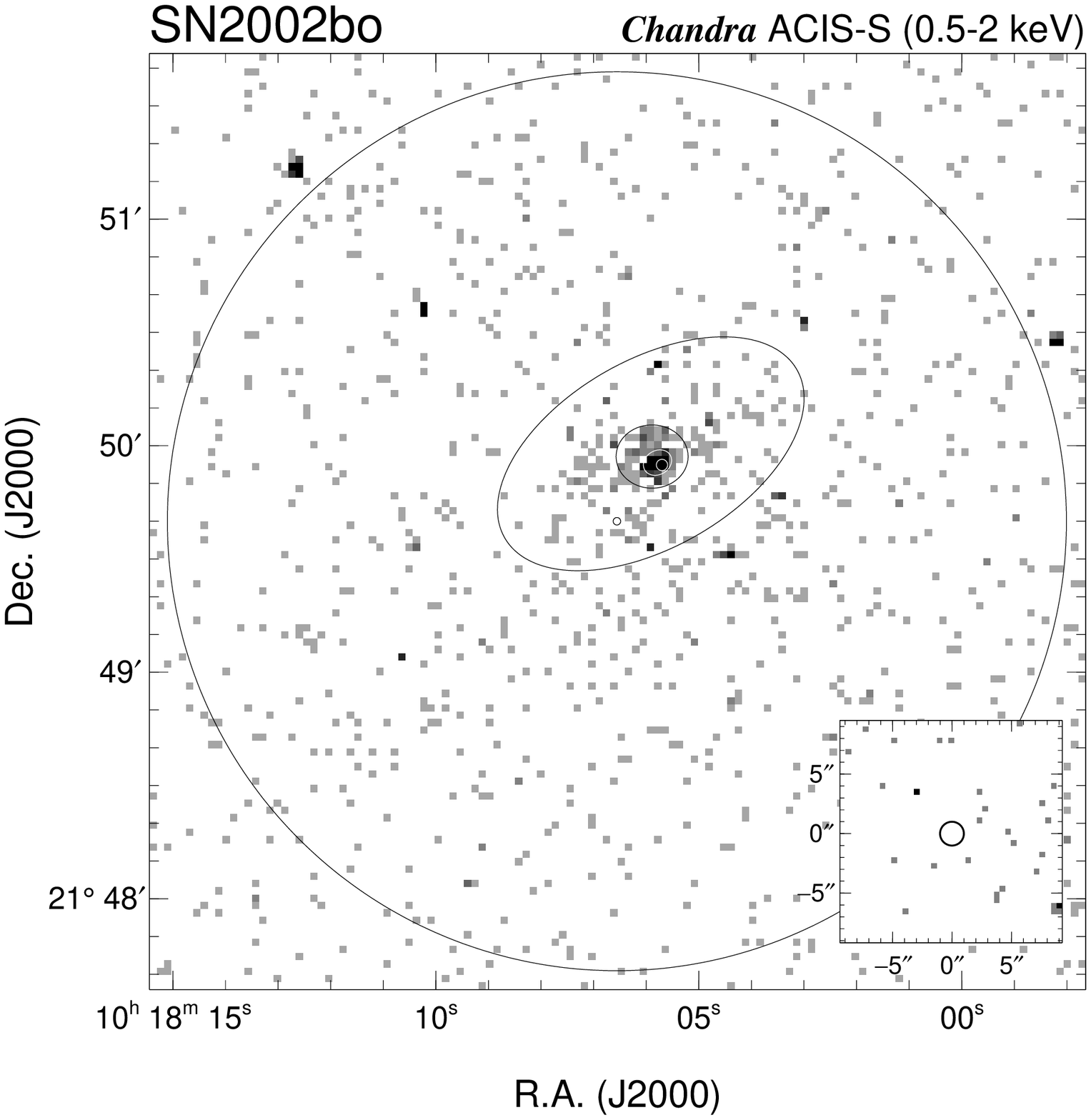}
\plotone{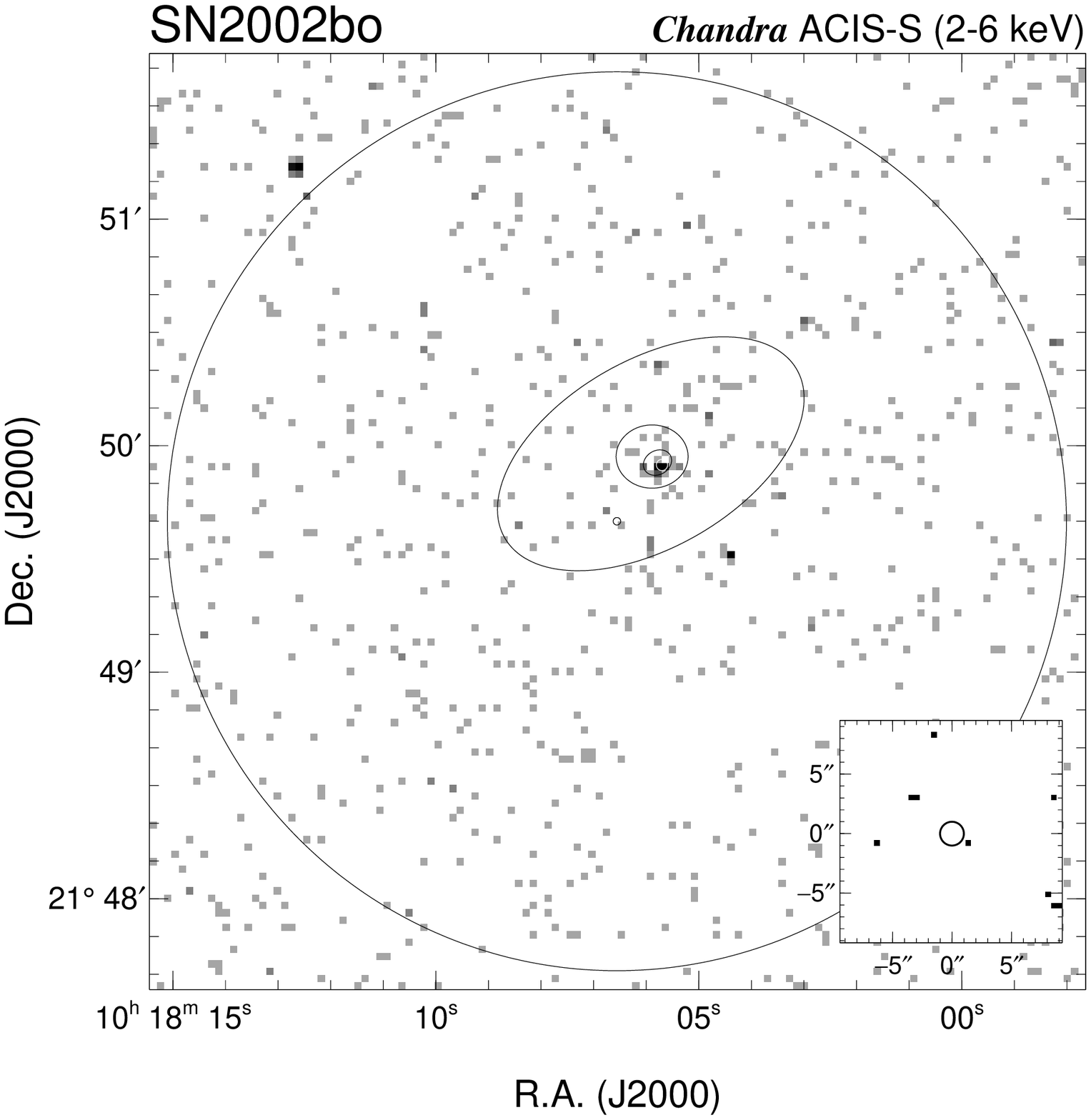}
\caption{
\chandra\ ACIS-S X-ray images of the vicinity of SN~2002bo
in the 0.5--2 keV (left) and 2--6 keV (right) bands displayed 
with 1.968$^{\prime\prime}$ pixel binning (block 4).  The various
circles and ellipses denote spectral extraction regions (see text).
The small insert panel at the bottom right shows the raw data
in the immediate vicinity of the SN in unblocked pixels
(i.e., 0.492$^{\prime\prime}$ $\times$ 0.492$^{\prime\prime}$). This
image is centered on the SN position and the axis labels are in
units of arcseconds as indicated.
}
\label{bo_soft_img}
\label{bo_hard_img}
\end{figure*}

\begin{figure*}
\epsscale{.45}
\plotone{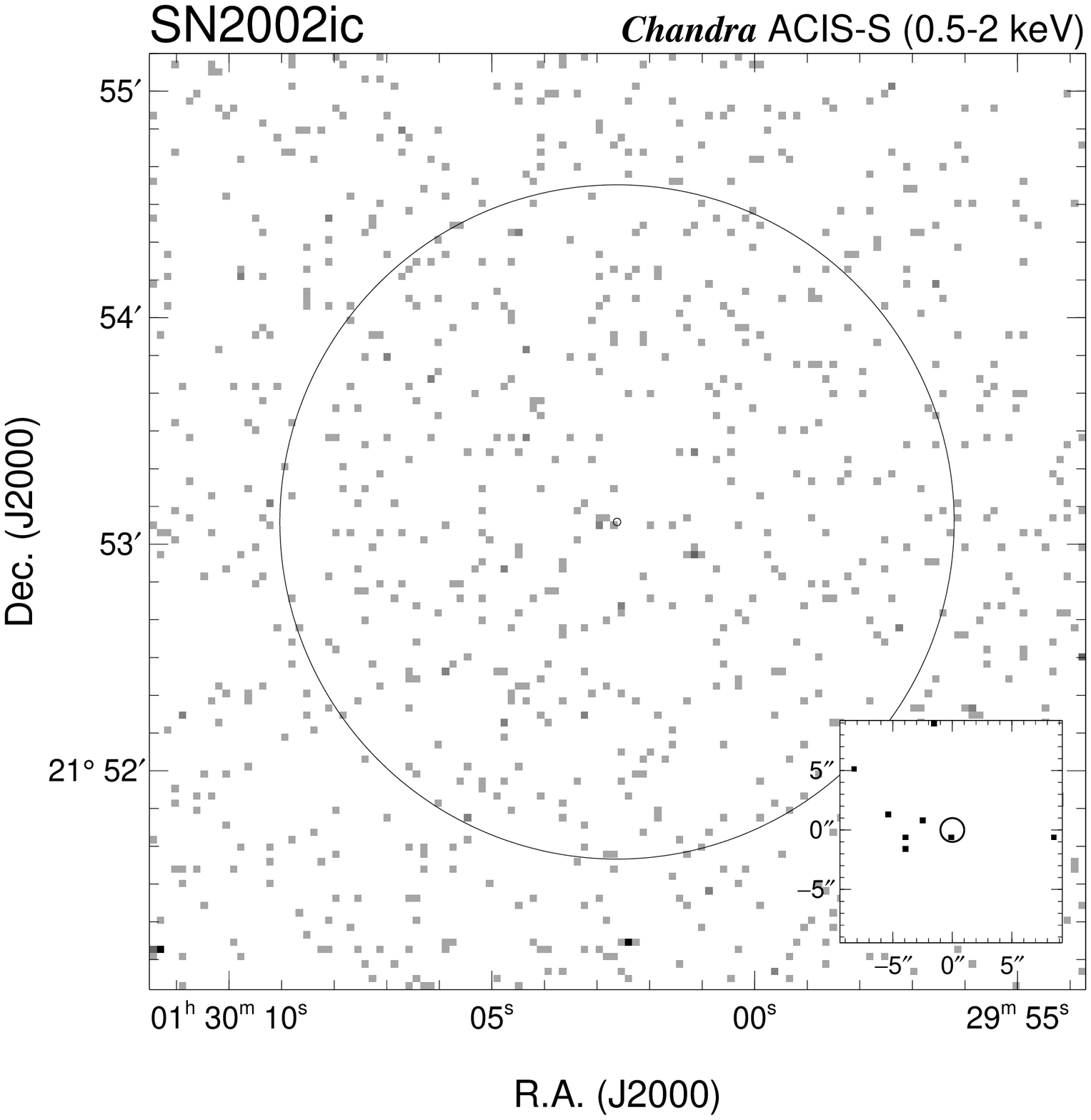}
\plotone{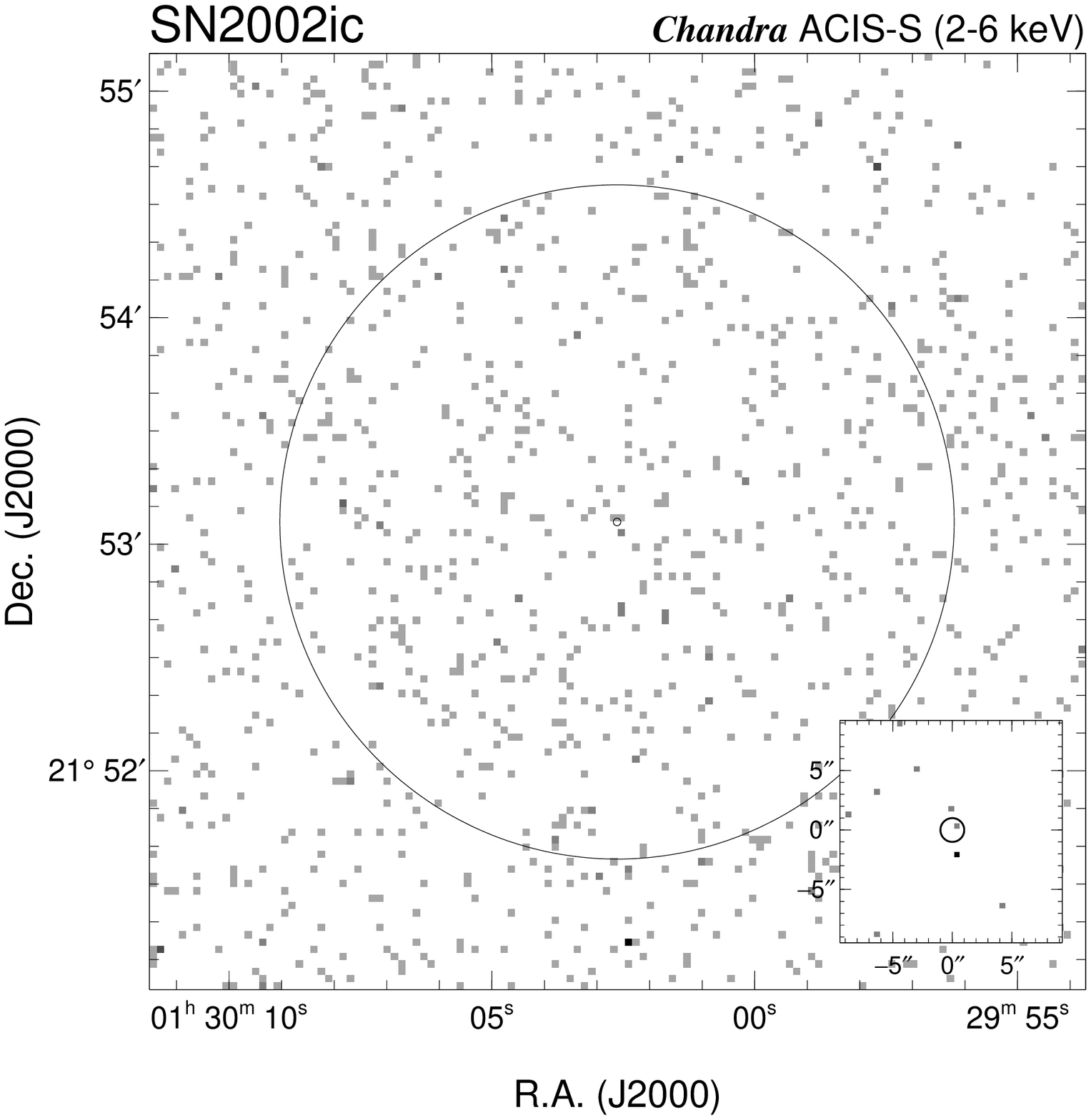}
\caption{ \chandra\ ACIS-S X-ray images of the vicinity of SN~2005ic.
The two circles denote spectral extraction regions (see text). Otherwise
similar to figure~\ref{bo_soft_img}.
}
\label{ic_soft_img}
\label{ic_hard_img}
\end{figure*}

\begin{figure*}
\epsscale{.45}
\plotone{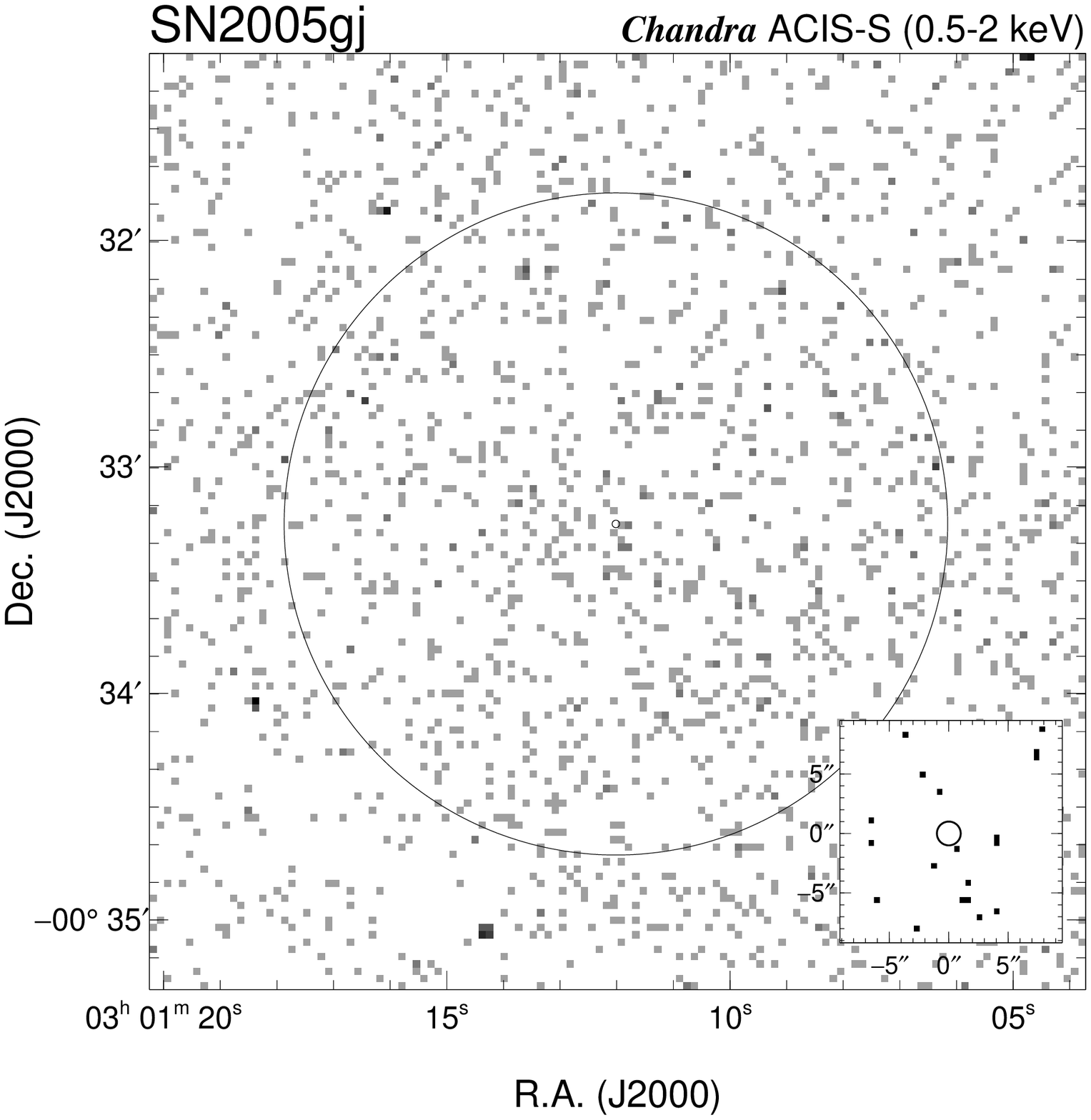}
\plotone{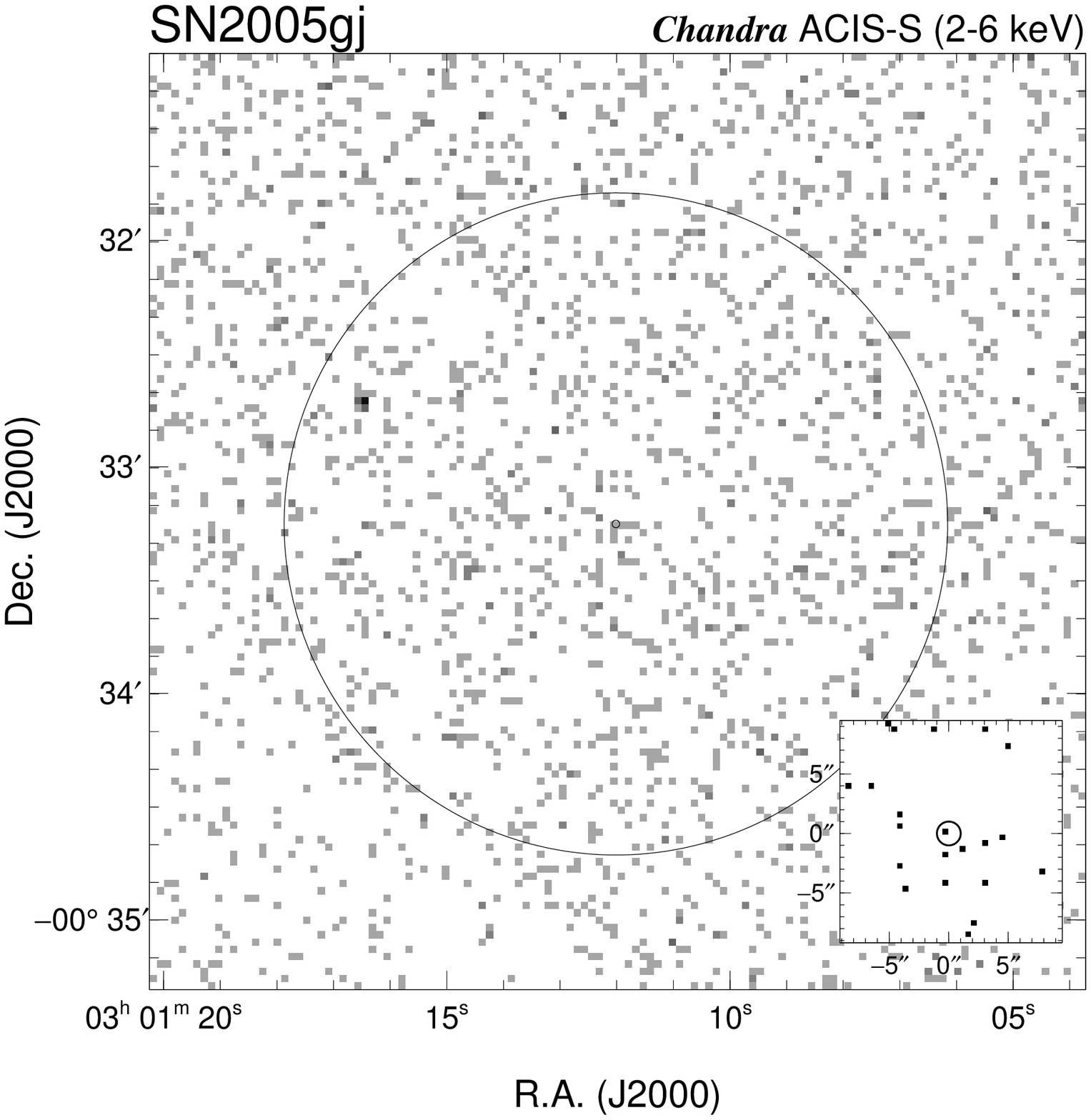}
\caption{ \chandra\ ACIS-S X-ray images of the vicinity of SN~2005gj.
Otherwise similar to figure~\ref{bo_soft_img}.
}
\label{gj_soft_img}
\label{gj_hard_img}
\end{figure*}

\begin{figure*}
\epsscale{.45}
\plotone{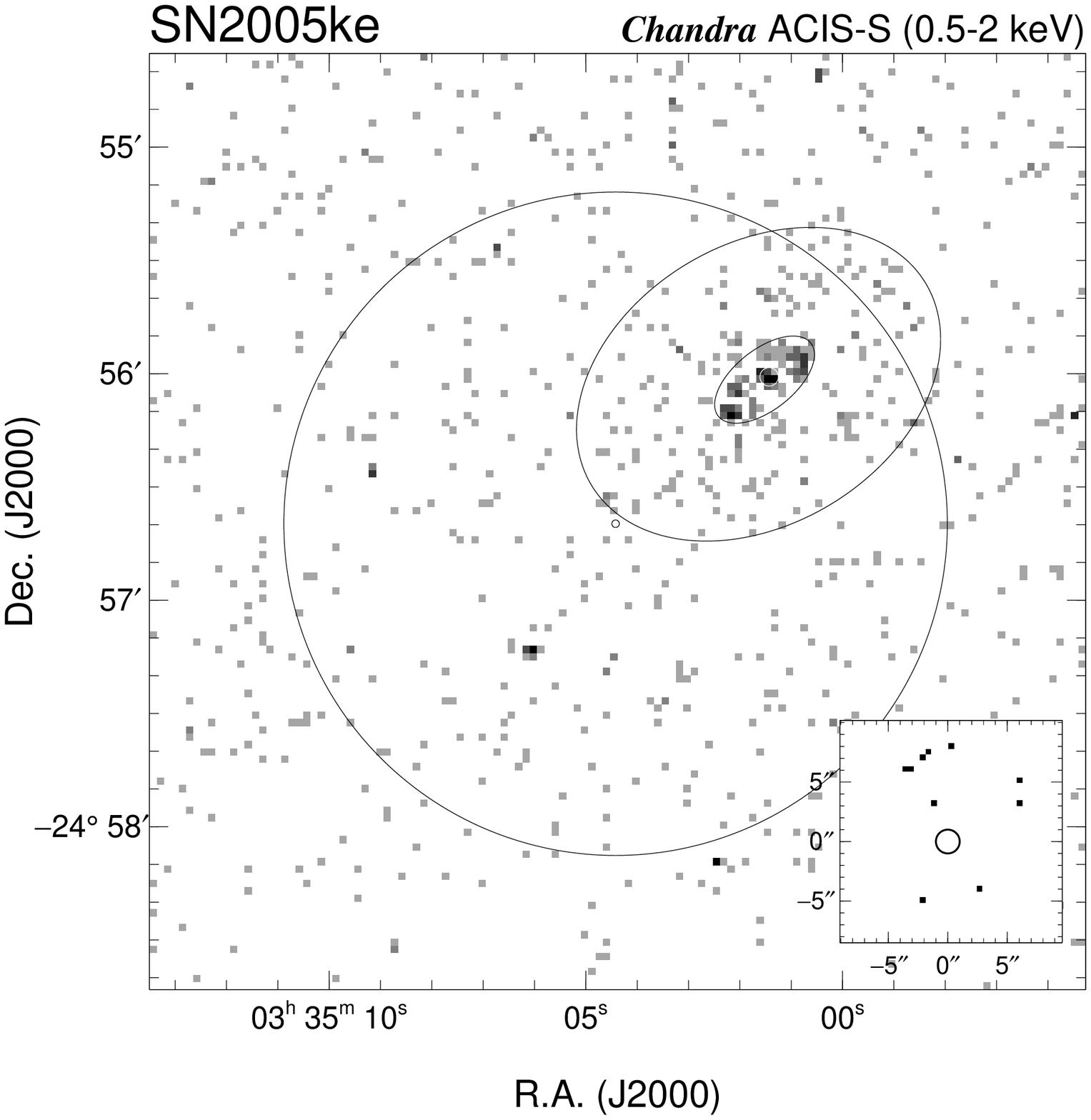}
\plotone{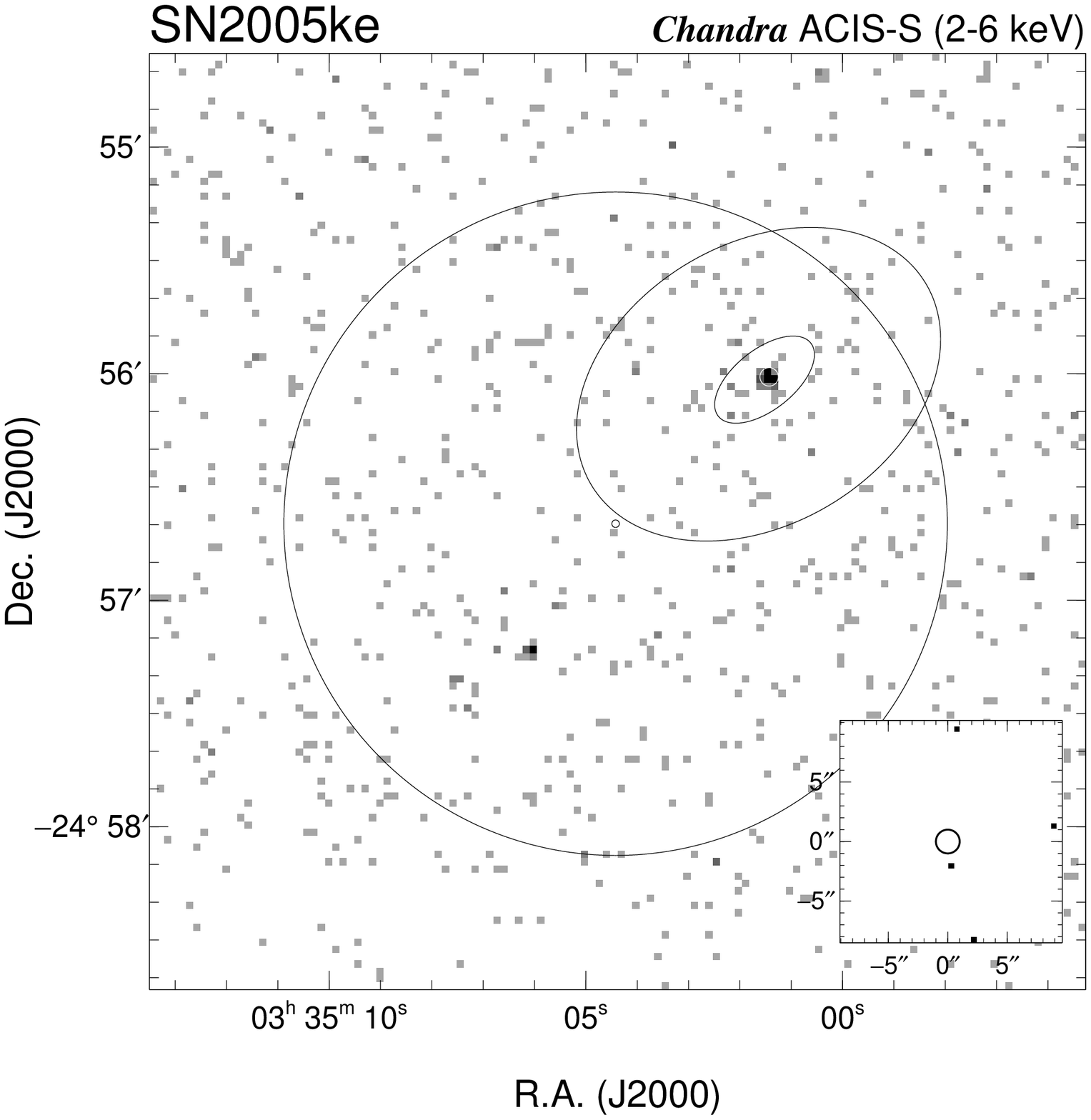}
\caption{ \chandra\ ACIS-S X-ray images of the vicinity of SN~2005ke.
The various circles and ellipses denote spectral extraction regions 
(see text). Otherwise similar to figure~\ref{bo_soft_img}.
}
\label{ke_soft_img}
\label{ke_hard_img}
\end{figure*}

Grayscale images of the \chandra\ data are presented in
Figures~\ref{bo_soft_img}--\ref{ke_soft_img} in two spectral bands:
0.5--2 keV (left) and 2-6 keV (right).  The small insert in the lower
right corner of each panel shows the raw \chandra\ data in the
immediate vicinity of each SN.  There are no detected X-ray photons in
a 1$^{\prime\prime}$ (radius) circle centered on the positions of
either SN~2002bo or SN~2005ke. Only two photons (with energies of
approximate 1.1 keV and 3.6 keV) were detected at the position of
SN~2002ic. Two photons (with energies of 4.0 keV and 7.7 keV) were also
detected at the position of SN~2002gj These are clearly not significant
detections and so in the following we demonstrate how we obtained
upper limits to the X-ray flux of each SN.  Since the range of allowed
spectral models is large, we developed a technique that allows us to
determine upper limits for just about any spectral form imaginable.

\begin{figure}
\epsscale{1.0}
\plotone{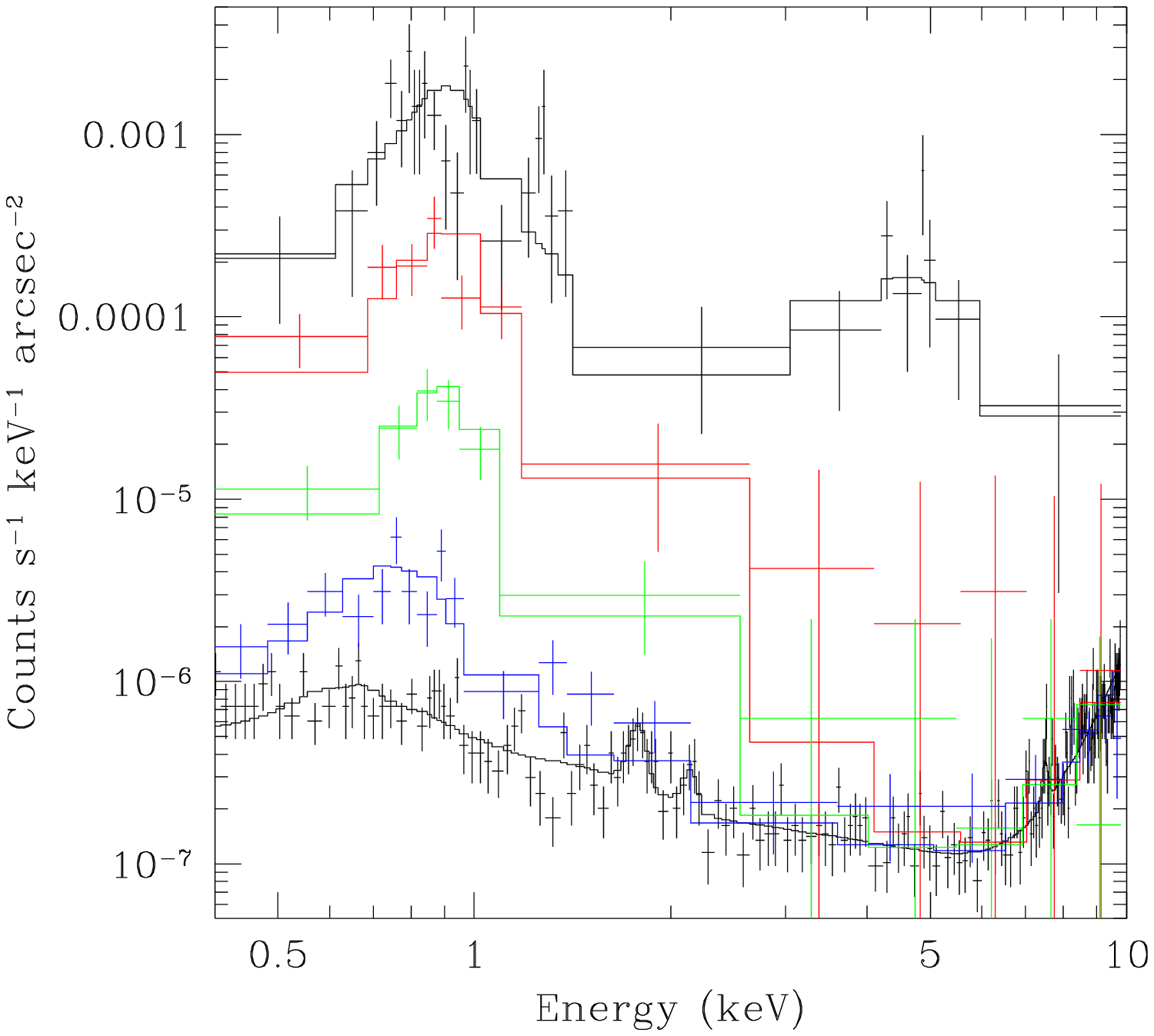}
\caption{ACIS-S spectra of the background (black, bottom) and 
several regions in NGC~3190: outer galaxy (blue), mid galaxy (green), 
inner galaxy (red), and nucleus (black, top).
}
\label{bo_spec}
\end{figure}

\begin{figure}
\epsscale{1.0}
\plotone{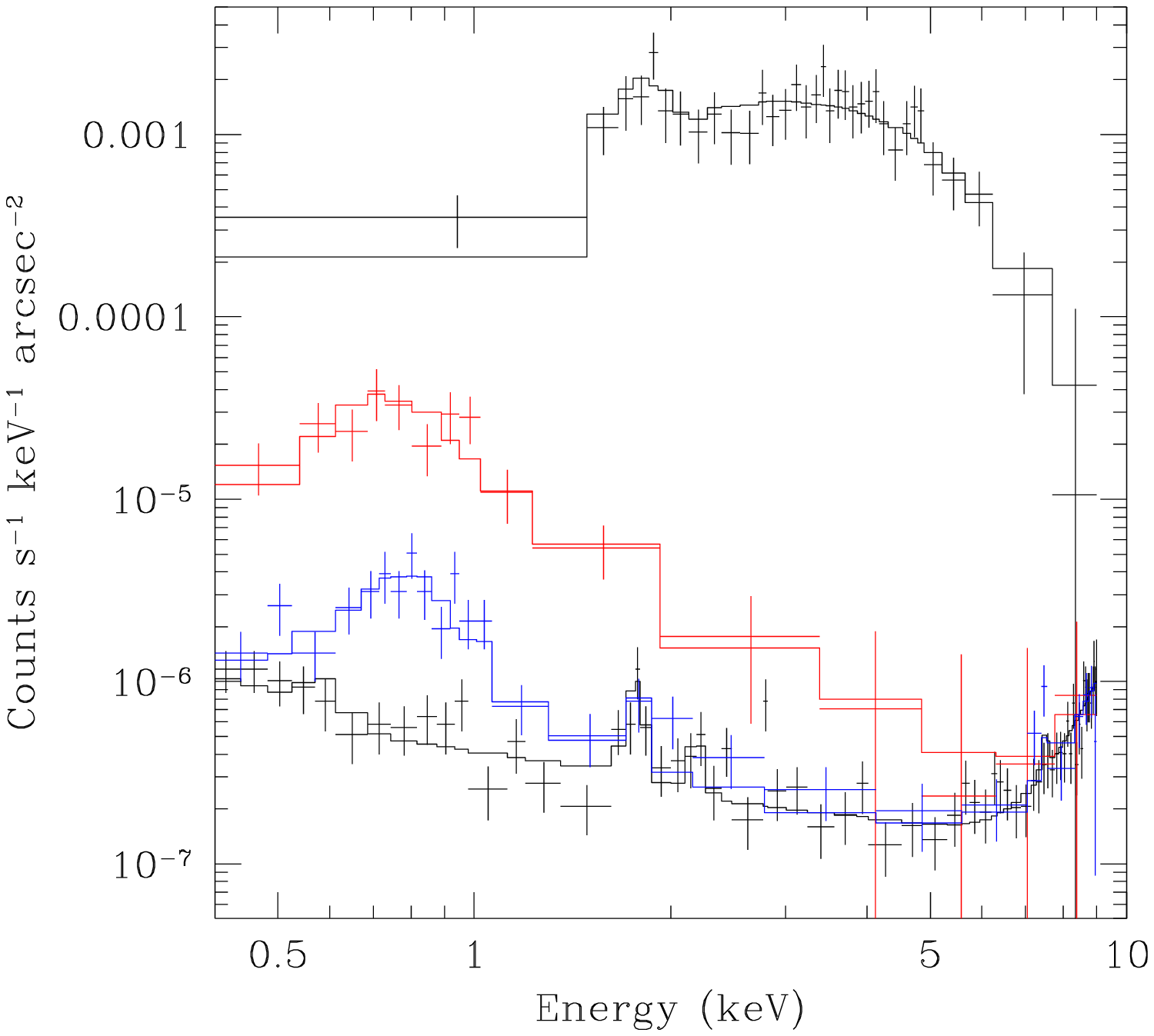}
\caption{ACIS-S spectra of the background (black, bottom) and several
regions in NGC~1371: outer galaxy (blue), inner galaxy (red), and
nucleus (black, top).  
}
\label{ke_spec}
\end{figure}

One essential key to our method is to obtain an accurate spectral
model for the background.  To this end we extracted spectra from the
large circular regions shown on
Figures~\ref{bo_soft_img}--\ref{ke_soft_img}, excluding all
serendipitous point sources and for each of SN~2002bo and SN~2002ke, an
elliptical region centered on the host galaxy that encompasses all the
galaxy's apparent X-ray emission. The detector spectral response
function (calculated using mkacisrmf) as well as the overall effective
area function were determined at a large number of detector positions
within the extraction region and then weighted by the spatial
distribution of detected events to produce final weighted response
functions. The black data points near the bottom of
Figures~\ref{bo_spec} and \ref{ke_spec} 
show the background spectra
used for SN~2002bo and SN~2005ke (the other cases are similar to
these two). Note that the spectra were rebinned to a
signal-to-noise ratio of $\sim$3 for display only; spectral fits were
carried out on the original unbinned spectral data.
The curve through these points indicates
our best fit model, which consists of two power law components to
describe the non-X-ray (i.e., particle) background, three narrow lines 
to account for instrumental
fluorescence lines from Si and Au, and two astrophysical 
models: a low
temperature thermal model for the local Galactic emission and a hard
power law for the unresolved cosmic X-ray background. These later
models included interstellar absorption using the Galactic values
\citep{dicloc90}: $N_{\rm H} = 2.1\times 10^{20}\, \rm cm^{-2}$
(SN~2002bo), $N_{\rm H} = 6.2\times 10^{20}\, \rm cm^{-2}$ (SN~2002ic),
$N_{\rm H} = 7.1\times 10^{20}\, \rm cm^{-2}$ (SN~2005gj), and $N_{\rm
H} = 1.4\times 10^{20}\, \rm cm^{-2}$ (SN~2005ke).  One of the
non-X-ray background power law models is required to model the
spectrum below $\sim$5 keV; the other one fits the sharp upturn in
emission beyond 7 keV.  Overall the non-X-ray background dominates the
spectrum above about 2 keV.  The fits were carried out using a maximum
likelihood figure-of-merit function for Poisson-distributed data (the
so-called ``c-stat'' option in xspec).  The background models for all
four targets are similar as are the quality of the fits, which are
acceptable from a statistical point-of-view (determined by examining the 
$\chi^2$ values on grouped data).

Detector spectral response and effective area functions were
calculated as above for the SN regions. We assume that the background
intensity is uniform on arcminute spatial scales near the detector
aimpoint and that its spectral form does not vary with position
either.  Thus we can scale the background models to the SN regions
using the ratio of detector pixels in the source and background
spectral extraction regions.  When this is done, we find that the
background model predicts the detection of 0.19 events (SN~2002bo),
0.20 events (SN~2002ic), 0.50 events (SN~2005gj), and 0.14 events
(SN~2005ke) over the entire Chandra band.  The number of events
actually detected in the observations of SN~2002bo and SN~2005ke (0 for
each) is consistent with their respective background rates. The Poisson
probability to detect 2 (or more) X-ray photons when 0.2 are expected,
as is the case for SN~2002ic, is only 1.7\%.  The Poisson probability
to detect 2 (or more) X-ray photons purely from background in the case
of SN~2005gj is 9\%.

\begin{figure*}
\epsscale{.4}
\plotone{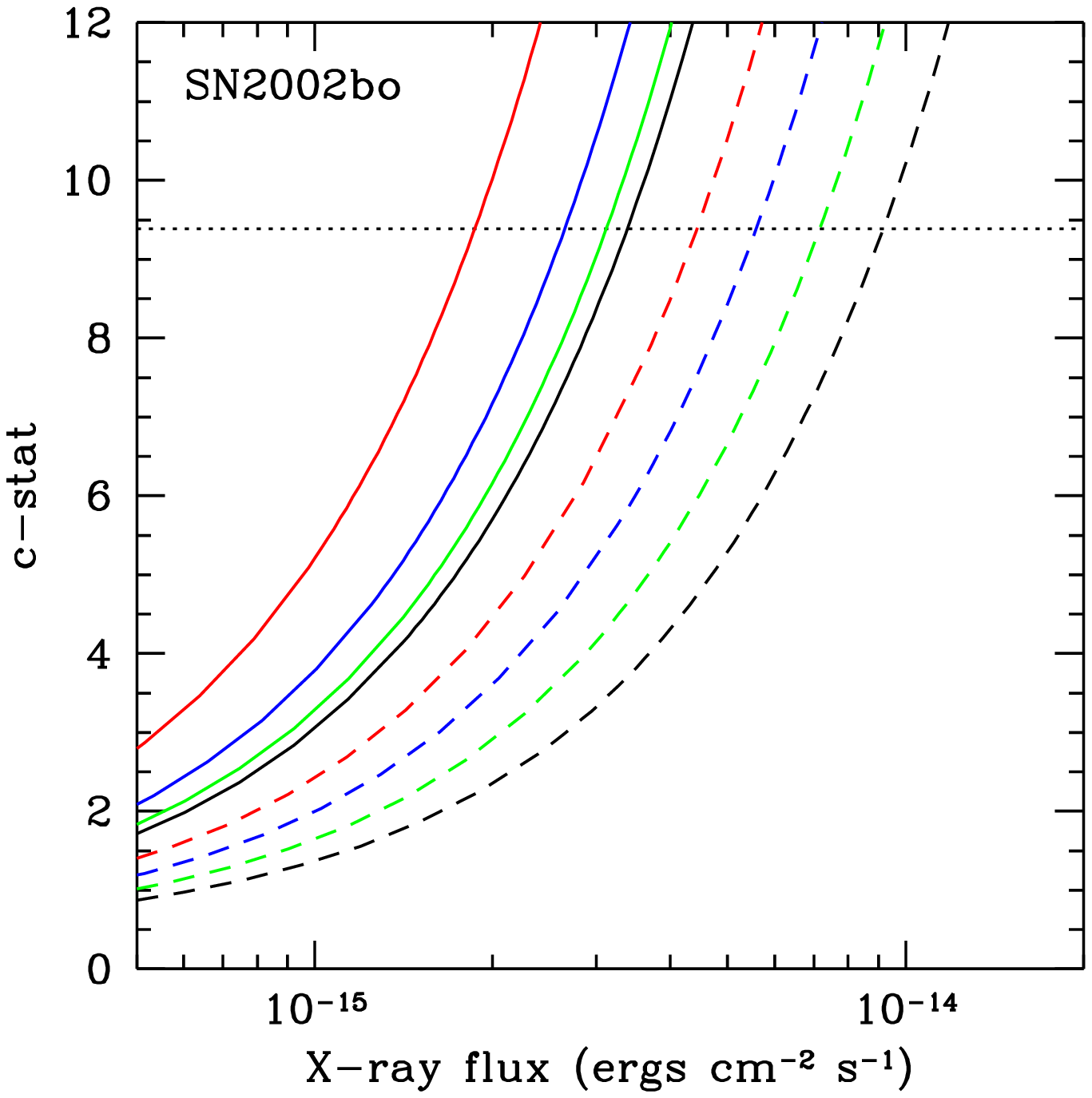}
\plotone{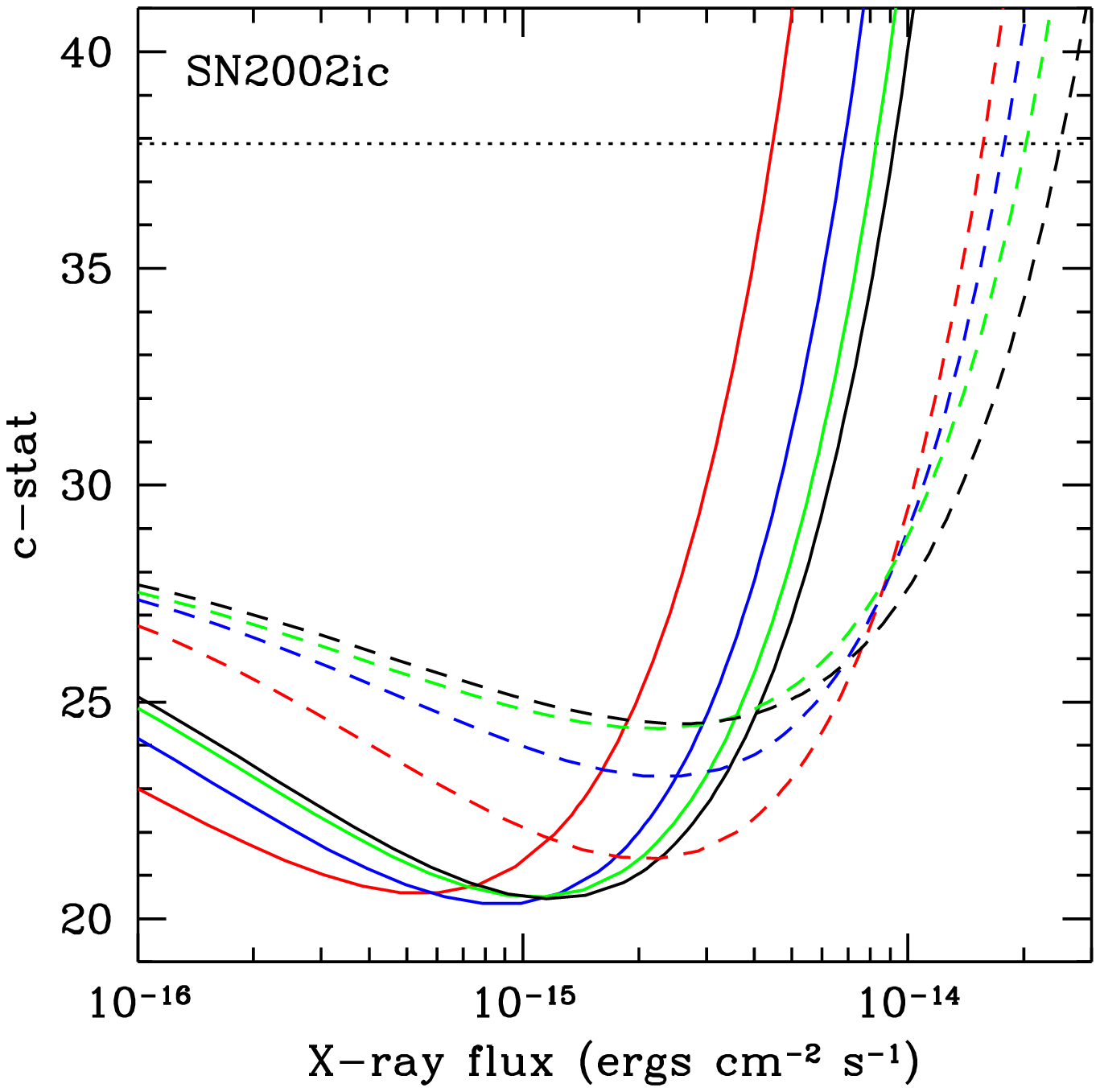}
\plotone{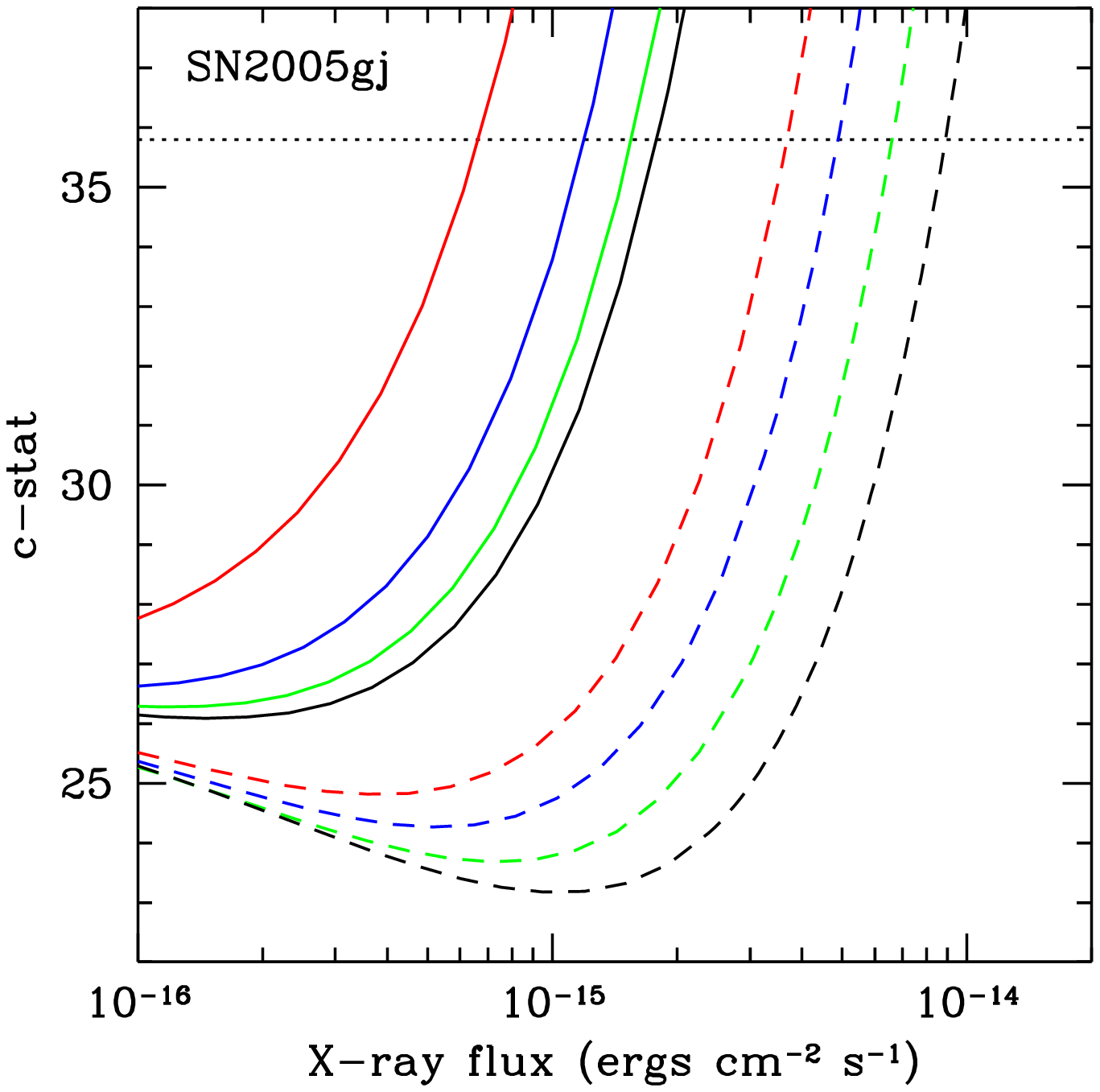}
\plotone{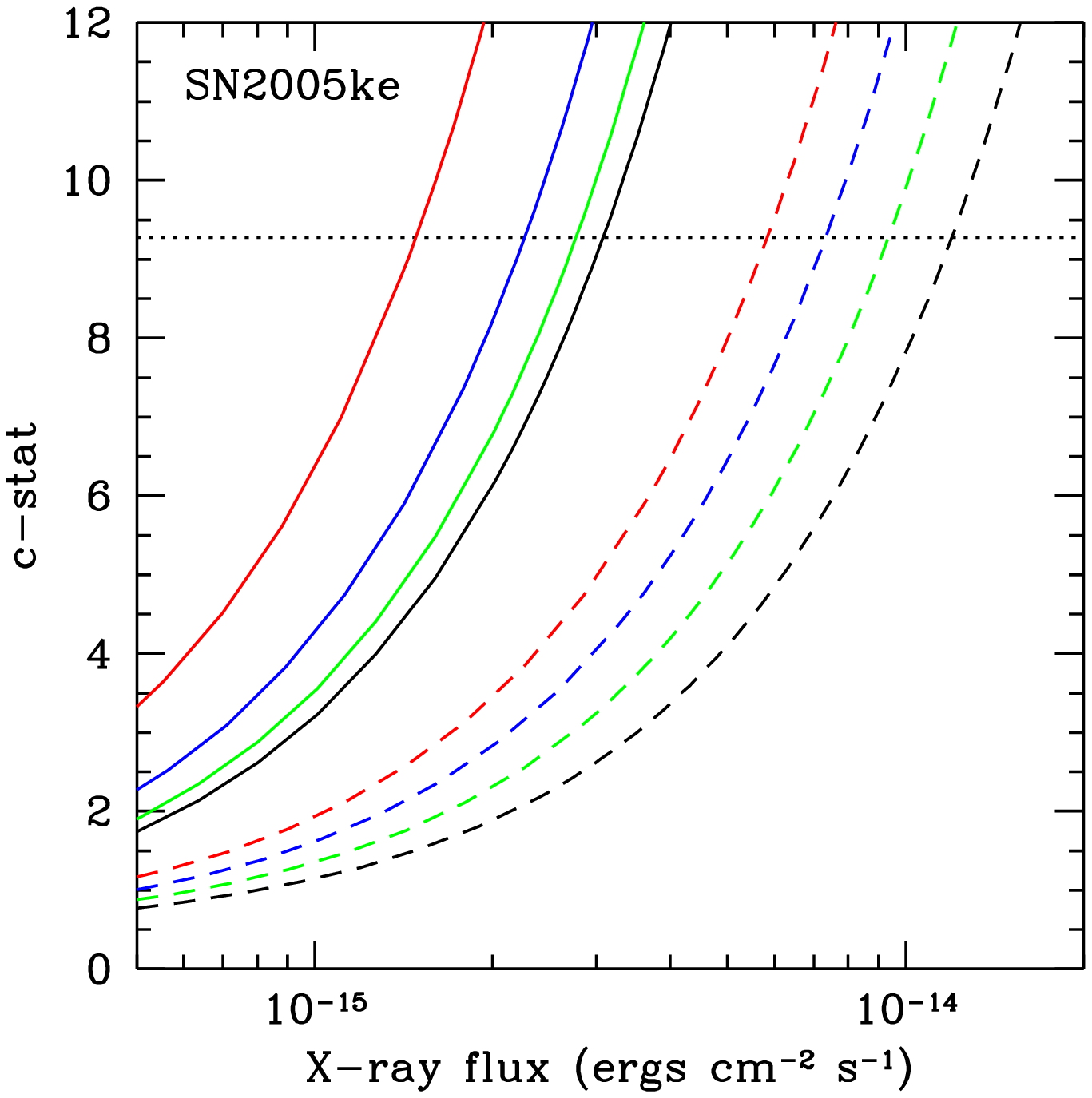}
\caption{ X-ray flux limits in the 0.5--10 keV band for SN~2002bo
(top left), SN~2002ic (top right), SN~2005gj (bottom left) and
SN~2005ke (bottom right).  The likelihood fit statistic
(``c-stat'') is plotted as a function of the X-ray flux limit. Solid
curves come from assuming thermal plasma models with different
temperatures: $kT=2$ (red), 5 (blue), 10 (green), and 20 (black) keV.
Dashed curves come from assuming a hot ($kT=80$ keV) bremsstrahlung
emission model with different amounts of intrinsic absorption: $N_{\rm
H} = 1\times 10^{22}$ (red), $N_{\rm H} = 2\times 10^{22}$ (blue),
$N_{\rm H} = 4\times 10^{22}$ (green), and $N_{\rm H} = 8\times 10^{22}$
(black) atoms cm$^{-2}$.  The horizontal dotted line near the top of
each panel denotes the 3-$\sigma$ criterion on the likelihood fit
statistic. The intersection of this line with the appropriate curve
provides the 3-$\sigma$ X-ray flux upper limit (to be read off the
x-axis).}
\label{xrfl}
\end{figure*}

We determine flux upper limits for a specific assumed astrophysical
spectral model (including the measured redshifts for SN~2002ic and
SN~2005gj), by adding it to the background model and then increasing
its normalization until the c-stat value increases by $\Delta = 9$,
which is the 3 $\sigma$ confidence level criterion.  Results for two
X-ray flux bands are shown in Tables~\ref{tab_bo} to \ref{tab_ke} and
plots of the variation of the c-stat value vs.~the total 0.5--10 keV
band X-ray flux are shown in Fig.~\ref{xrfl}.  All quoted fluxes have
been corrected to remove Galactic absorption for SN~2002ic, SN~2005gj,
and SN~2005ke, and the absorption from the host galaxy for
SN~2002bo. The first set of four models in these tables (dashed curves
in the figure) assume thermal plasma emission with no intrinsic
absorption, normal solar composition and temperatures from 2 to 20
keV.  The second set of four models (solid curves in the figure)
assume a hot ($kT=80$ keV) bremsstrahlung model with varying amounts
of intrinsic absorption: from $N_{\rm H}= 1\times 10^{22}\,\rm
cm^{-2}$ to $N_{\rm H}=8\times 10^{22}\,\rm cm^{-2}$. Uncertainty in
the level of the background has a very small effect on the flux upper
limits.  For example, assuming a 20\% variation in the level of the
model background for SN~2002ic introduces a $\sim$3\% change in
the derived flux limit for the hard spectral model.

\begin{deluxetable}{cccc}
\tablewidth{0pt}
\tablecaption{X-ray Flux Limits (3 $\sigma$) for SN~2002bo}
\tablecolumns{4}
\tablehead{
  \colhead{$kT$} & 
  \colhead{Intrinsic $N_{\rm H}$} &
  \colhead{Flux} &
  \colhead{Flux} \\
  \colhead{(keV)} & 
  \colhead{(atoms cm$^{-2}$)} &
  \colhead{(0.5--2 keV)} &
  \colhead{(2--10 keV)} }
\startdata
 2  & $\dots$  &  $1.2\times 10^{-15}$  &  $6.8\times 10^{-16}$ \\
 5  & $\dots$  &  $9.9\times 10^{-16}$  &  $1.7\times 10^{-15}$ \\
 10 & $\dots$  &  $9.0\times 10^{-16}$  &  $2.2\times 10^{-15}$ \\
 20 & $\dots$  &  $8.5\times 10^{-16}$  &  $2.5\times 10^{-15}$ \\
 80 & $1\times 10^{22}$   &  $3.2\times 10^{-16}$  &  $4.1\times 10^{-15}$ \\
 80 & $2\times 10^{22}$   &  $2.0\times 10^{-16}$  &  $5.4\times 10^{-15}$ \\
 80 & $4\times 10^{22}$   &  $8.3\times 10^{-17}$  &  $7.1\times 10^{-15}$ \\
 80 & $8\times 10^{22}$   &  $1.4\times 10^{-17}$  &  $9.2\times 10^{-15}$ \\
\enddata
\label{tab_bo}
\end{deluxetable}

\begin{deluxetable}{cccc}
\tablewidth{0pt}
\tablecaption{X-ray Flux Limits (3 $\sigma$) for SN~2002ic}
\tablecolumns{4}
\tablehead{
  \colhead{$kT$} & 
  \colhead{Intrinsic $N_{\rm H}$} &
  \colhead{Flux} &
  \colhead{Flux} \\
  \colhead{(keV)} & 
  \colhead{(atoms cm$^{-2}$)} &
  \colhead{(0.5--2 keV)} &
  \colhead{(2--10 keV)} }
\startdata
 2  & $\dots$  &  $3.0\times 10^{-15}$  &  $1.5\times 10^{-15}$ \\
 5  & $\dots$  &  $2.7\times 10^{-15}$  &  $4.2\times 10^{-15}$ \\
 10 & $\dots$  &  $2.5\times 10^{-15}$  &  $5.8\times 10^{-15}$ \\
 20 & $\dots$  &  $2.4\times 10^{-15}$  &  $6.9\times 10^{-15}$ \\
 80 & $1\times 10^{22}$    &  $1.1\times 10^{-15}$  &  $1.5\times 10^{-14}$ \\
 80 & $2\times 10^{22}$    &  $6.5\times 10^{-16}$  &  $1.7\times 10^{-14}$ \\
 80 & $4\times 10^{22}$    &  $2.4\times 10^{-16}$  &  $2.0\times 10^{-14}$ \\
 80 & $8\times 10^{22}$    &  $3.8\times 10^{-17}$  &  $2.5\times 10^{-14}$ \\
\multispan2{Chugai Model -- Day 260}          &  $9.0\times 10^{-17}$  &  $2.2\times 10^{-14}$ \\
\enddata
\label{tab_ic}
\end{deluxetable}

\begin{deluxetable}{cccc}
\tablewidth{0pt}
\tablecaption{X-ray Flux Limits (3 $\sigma$) for SN~2005gj}
\tablecolumns{4}
\tablehead{
  \colhead{$kT$} & 
  \colhead{Intrinsic $N_{\rm H}$} &
  \colhead{Flux} &
  \colhead{Flux} \\
  \colhead{(keV)} & 
  \colhead{(atoms cm$^{-2}$)} &
  \colhead{(0.5--2 keV)} &
  \colhead{(2--10 keV)} }
\startdata
 2  & $\dots$           &  $4.4\times 10^{-16}$  &  $2.3\times 10^{-16}$ \\
 5  & $\dots$           &  $4.7\times 10^{-16}$  &  $7.3\times 10^{-16}$ \\
 10 & $\dots$           &  $4.6\times 10^{-16}$  &  $1.1\times 10^{-15}$ \\
 20 & $\dots$           &  $4.6\times 10^{-16}$  &  $1.3\times 10^{-15}$ \\
 80 & $1\times 10^{22}$ &  $2.7\times 10^{-16}$  &  $3.4\times 10^{-15}$ \\
 80 & $2\times 10^{22}$ &  $1.8\times 10^{-16}$  &  $4.7\times 10^{-15}$ \\
 80 & $4\times 10^{22}$ &  $7.7\times 10^{-17}$  &  $6.5\times 10^{-15}$ \\
 80 & $8\times 10^{22}$ &  $1.4\times 10^{-17}$  &  $8.9\times 10^{-15}$ \\
\enddata
\label{tab_gj}
\end{deluxetable}

\begin{deluxetable}{cccc}
\tablewidth{0pt}
\tablecaption{X-ray Flux Limits (3 $\sigma$) for SN~2005ke}
\tablecolumns{4}
\tablehead{
  \colhead{$kT$} & 
  \colhead{Intrinsic $N_{\rm H}$} &
  \colhead{Flux} &
  \colhead{Flux} \\
  \colhead{(keV)} & 
  \colhead{(atoms cm$^{-2}$)} &
  \colhead{(0.5--2 keV)} &
  \colhead{(2--10 keV)} }
\startdata
 2  & $\dots$           &  $9.4\times 10^{-16}$  &  $5.4\times 10^{-16}$ \\
 5  & $\dots$           &  $8.5\times 10^{-16}$  &  $1.4\times 10^{-15}$ \\
 10 & $\dots$           &  $8.0\times 10^{-16}$  &  $2.0\times 10^{-15}$ \\
 20 & $\dots$           &  $7.7\times 10^{-16}$  &  $2.3\times 10^{-15}$ \\
 80 & $1\times 10^{22}$ &  $4.2\times 10^{-16}$  &  $5.4\times 10^{-15}$ \\
 80 & $2\times 10^{22}$ &  $2.7\times 10^{-16}$  &  $7.1\times 10^{-15}$ \\
 80 & $4\times 10^{22}$ &  $1.1\times 10^{-16}$  &  $9.2\times 10^{-15}$ \\
 80 & $8\times 10^{22}$ &  $1.8\times 10^{-17}$  &  $1.2\times 10^{-14}$ \\
\enddata
\label{tab_ke}
\end{deluxetable}

\begin{figure}
\epsscale{1.0}
\plotone{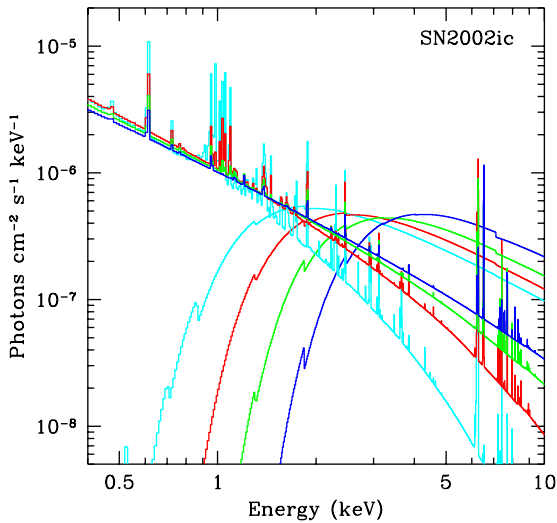}
\caption{Maximum allowed X-ray emission (at 3 $\sigma$)
from SN~2002ic for several different assumed spectral forms. 
Curves with emission lines are for thermal models with 
normal (solar) composition,  no intrinsic absorption, and a 
range of different temperatures: 
$kT=2$ keV (cyan), $kT=5$ keV (red), $kT=10$ keV (green), $kT=20$ keV 
(blue).
Smooth curves are for a hard bremsstrahlung model ($kT=80$ keV) and
varying amounts of intrinsic absorption: $N_{\rm H}= 1\times 10^{22}\,\rm
cm^{-2}$ (cyan), $N_{\rm H}= 2\times 10^{22}\,\rm cm^{-2}$ (red),
$N_{\rm H}= 4\times 10^{22}\,\rm cm^{-2}$ (green), and $N_{\rm H}=
8\times 10^{22}\,\rm cm^{-2}$ (blue).}
\label{ic_speclimits}
\end{figure}

Fig.~\ref{ic_speclimits} graphically illustrates how the upper
limits depend on the assumed spectral shape.  
Each curve here corresponds
to the upper limit for that spectrum allowed by the \chandra\ data.
Clearly more hard band emission is allowed if the soft band is
depressed, and vice versa.

Some comments on Fig.~\ref{xrfl} are warranted.  The top left panel
(SN~2002bo) shows a steady rise in the c-stat value as the flux (or
normalization) of the assumed spectral model is increased. The top
right panel (SN~2002ic) shows a slightly different behavior: the c-stat
value initially {\it decreases} as the flux of the model increases
until it reaches a minimum after which it begins to grow rapidly.
What is happening is that the added spectral model is accounting for
the 2 detected photons in the SN~2002ic spectrum.  The reduction in the
c-stat value is $\Delta \sim 8$ which is significant at between 2 and
3 $\sigma$.  The curves for SN~2005gj resemble those for SN~2002ic for
essentially the same reason, while the curves for SN~2005ke (for which
no photons were detected) resemble SN~2002bo.

Finally we compare our X-ray flux limits for SN~2005gj with those
quoted by \citet{prietoetal07}: $5\times 10^{-16}$ ergs cm$^{-2}$
s$^{-1}$ (at 68\% confidence level) and $9\times 10^{-16}$ ergs
cm$^{-2}$ s$^{-1}$ (at 95.5\% confidence level) for the 0.5--8 keV
band.  The spectral form they assumed (an unabsorbed power law with
photon index $\Gamma = -2$) is roughly comparable to our thermal
plasma model with $kT = 10$ keV.  We utilize Fig.~\ref{xrfl} to
convert to the 68\% (c-stat $\Delta = 1$) and 95.5\% (c-stat $\Delta =
4$) confidence levels and obtain comparable flux limits (0.5--10 keV
band) of $4\times 10^{-16}$ ergs cm$^{-2}$ s$^{-1}$ and $9\times
10^{-16}$ ergs cm$^{-2}$ s$^{-1}$, respectively.

\subsection{Re-evaluation of tentative Swift X-ray detection of SN~2005ke}

The importance of the tentative detection of SN~2005ke by the Swift
X-ray Telescope (XRT) instrument cannot be overstated, yet the
original report \citep{immler06} provided only a brief summary of the
data reduction and analysis.  We undertook an independent analysis of
these data in order to address some issues that we considered to be
incompletely resolved in the published study.

We downloaded the archival Swift XRT data on SN~2005ke observed during
the winter of 2005/2006 from the HEASARC. There were 43 separate
observations from 14 November 2005 until 5 March 2006 encompassing
ObsIDs from 0030341001 to 0030341008 and from 0030342001 to
0030342036. Recently (March 2007) Swift reobserved SN~2005ke (ObsIDs
0035898002 to 0035898004) and we examined these data as well.  The
report on this recent observation can be summarized succinctly: no
photons were detected within 14$^{\prime\prime}$ of the SN position.
In our analysis of the rest of the data we began with the standard
cleaned events files utilizing only data in photon counting mode and
all good grades (0-12).

The position of the nuclear X-ray source was located in each of the
separate observations and then images in a broad band (0.5-6 keV, pi
bins 50:600) were generated using this location for the image center.
The four shortest observations (exposures $<$ 100 s) gave no detected
photons from the nucleus and were excluded from further consideration.
The total dead-time corrected exposure in the merged image was 251.0
ks.

\begin{figure*}
\epsscale{.375}
\plotone{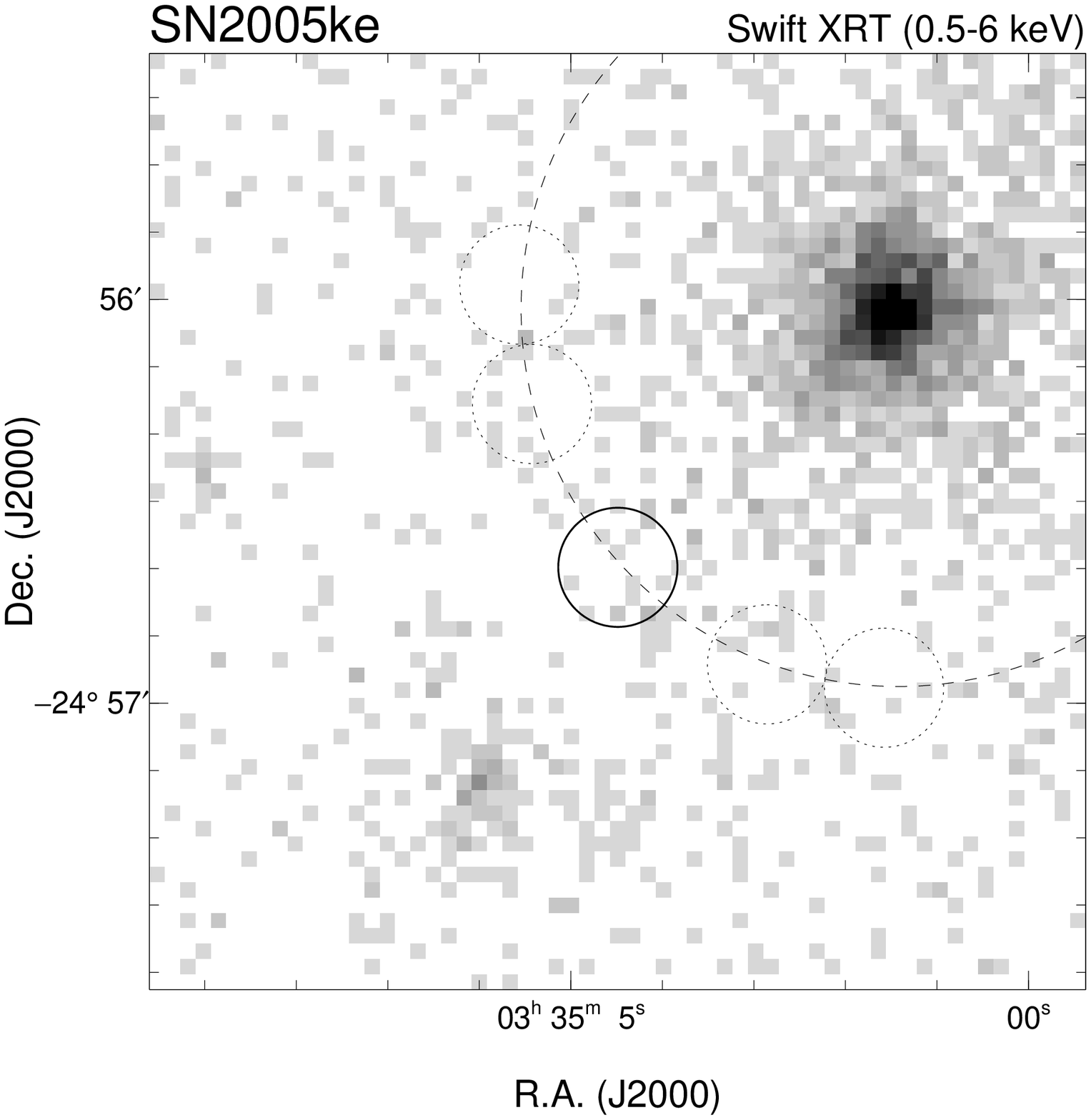}
\plotone{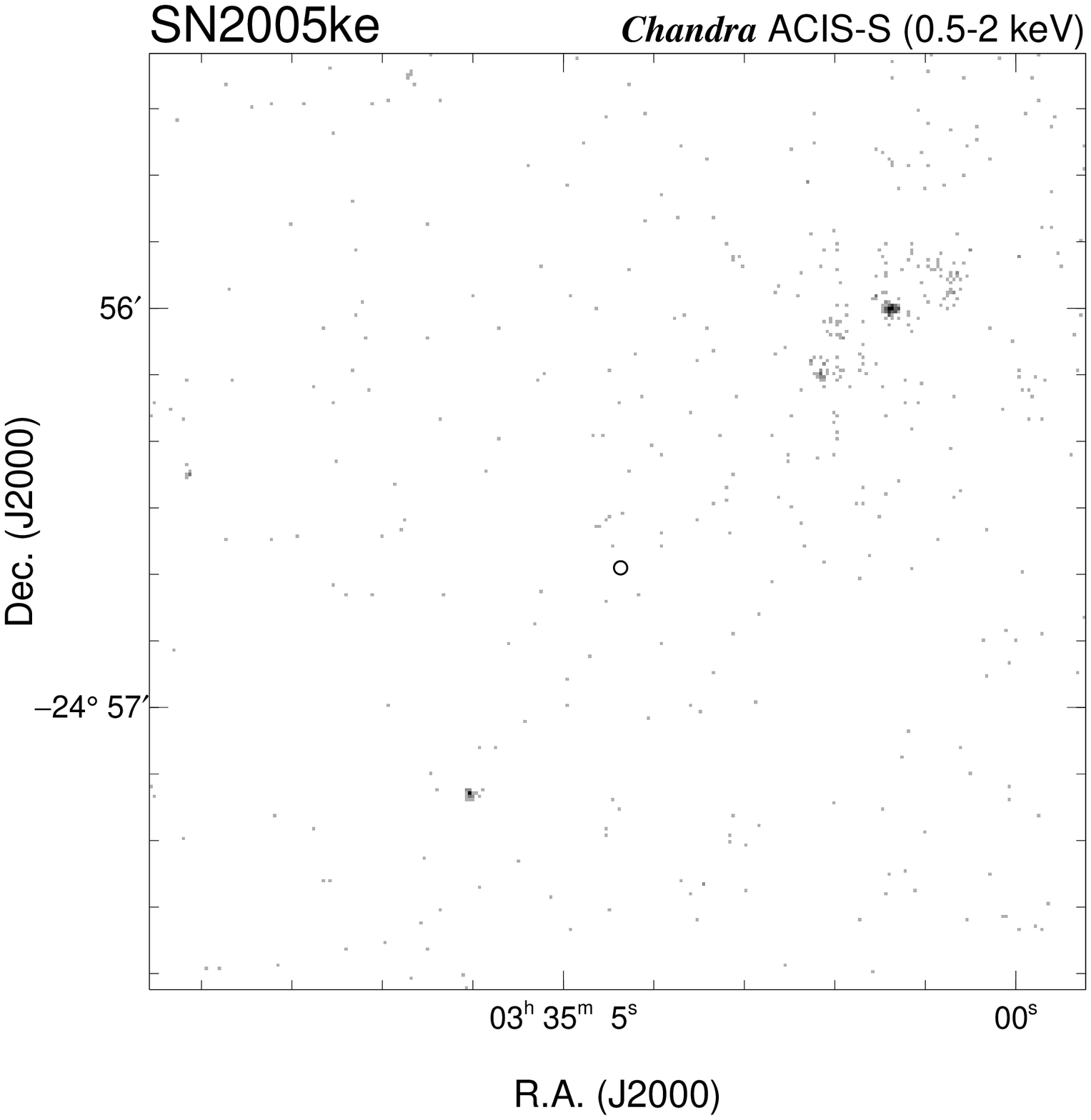}
\plotone{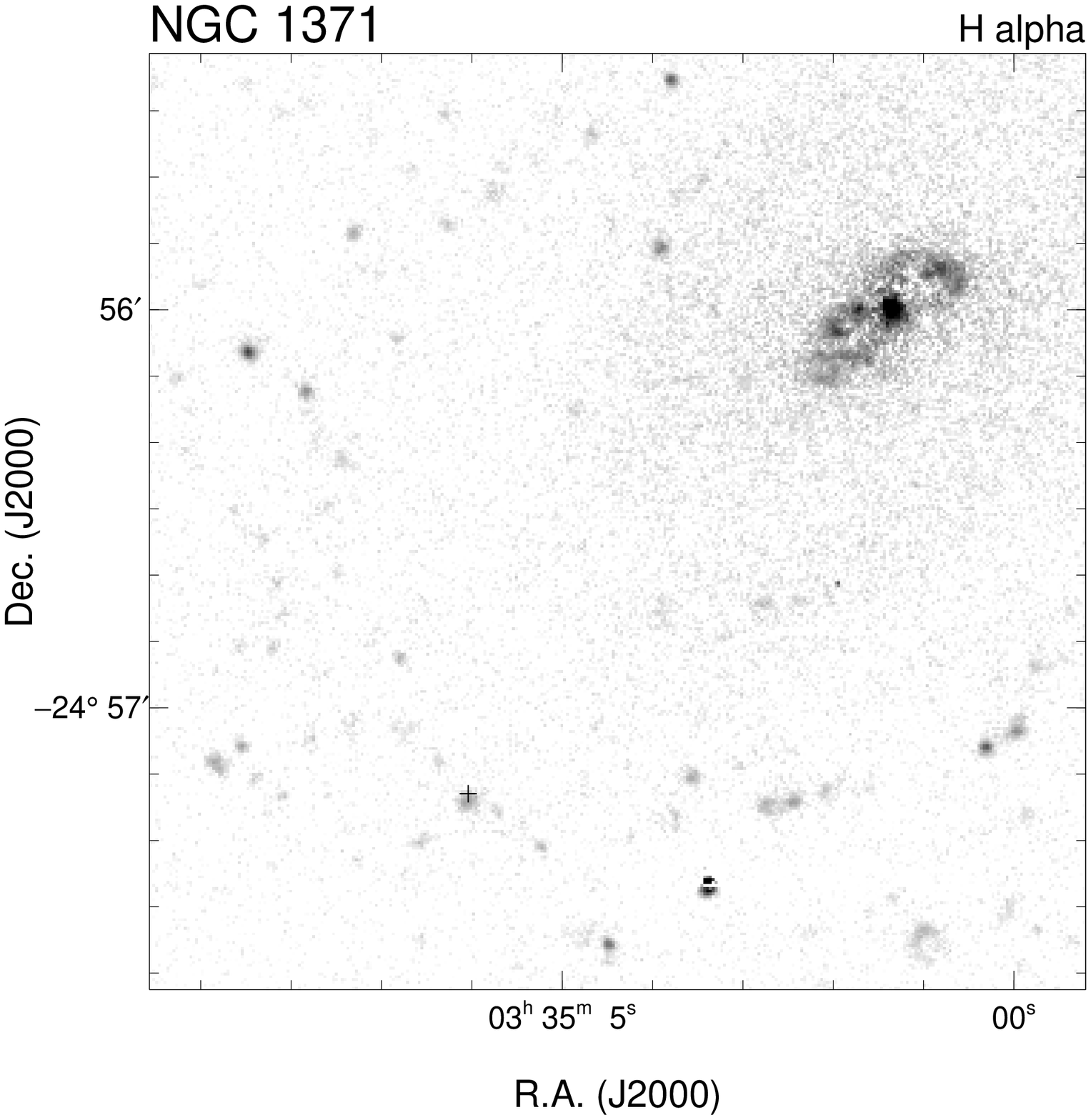}
\caption{
The left and middle panels show X-ray images of the vicinity of
SN~2005ke from the Swift XRT (left) and \chandra\ ACIS-S (middle).
The solid-linetype circle near the middle of these panels denote the
extraction regions for SN~2005ke (9$^{\prime\prime}$ radius for Swift,
1$^{\prime\prime}$ radius for \chandra).  In the case of Swift the
small dotted-linetype circles denote the locations from which
background was obtained.  Note that the 4 background and single source 
regions are equidistant
from the center of the bright nuclear emission (as noted by the large
dashed-linetype circle which has a 1$^\prime$ radius).
The right panel shows the H$\alpha$ emission
from NCG~1371 originally presented in \citet{hamdev99}. 
All three images cover the same field of view.
}
\label{ke_swchha_img}
\end{figure*}

We registered the Swift data to the \chandra\ data using the observed
positions of 11 fairly bright X-ray sources that contained at least 30
counts in the merged Swift image.  A simple linear 4$^{\prime\prime}$
shift in position was sufficient to bring the Swift and \chandra\
positions into agreement with a scatter in the relative after-shift
positions of $<$1$^{\prime\prime}$.  
The accuracy of the \chandra\ positional
astrometry was verified, as mentioned above, using the cataloged
position of an optical star detected in the X-rays.
Figure~\ref{ke_swchha_img} shows the Swift (left panel) and \chandra\
(middle panel) broad band images highlighting the region covering the
supernova position and the nucleus of NGC~1371.

\begin{figure}
\epsscale{1.0}
\plotone{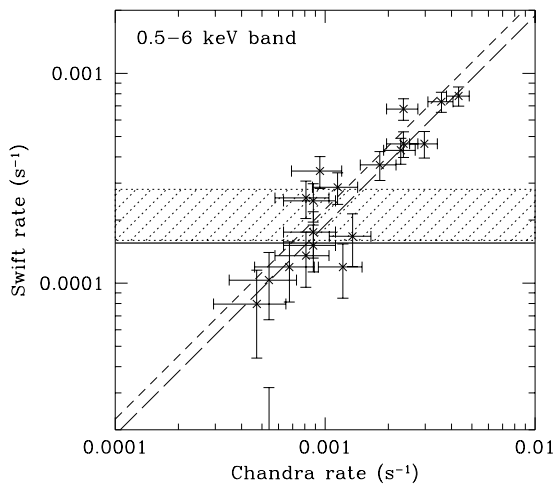}
\caption{
\chandra\ ACIS-S vs.\ Swift XRT count rates and 1 $\sigma$ error bars
for unresolved X-ray sources in the field of view of the SN~2005ke
observations.  The long dashed line shows the mean ratio of rates from
the measured data points.  This is in good agreement with the ratio of
effective areas at 1.5 keV, which is shown as the short dashed line.
The hatched region across the middle of the plot shows the claimed
detection count rate and 1 $\sigma$ error from \citet{immler06}, while
the thick dark line shows our 3 $\sigma$ upper limit to the count rate.
}
\label{ke_compare_rates}
\end{figure}

The broadband Swift and \chandra\ count rates were determined for the
aforementioned 11 sources in addition to 9 other fainter ones. These
are essentially all the compact X-ray sources within $\sim$5$^\prime$
of the \chandra\ optical axis.  For Swift the rates were extracted
from within 9$^{\prime\prime}$ radius circular regions and then
multiplied by two to account for the point-spread function (PSF) of
the telescope (the half-power diameter at 1.5 keV is
18$^{\prime\prime}$). The \chandra\ point source extraction regions
were large enough to capture $>$90\% of the source flux.
Figure~\ref{ke_compare_rates} plots the Swift vs.\ \chandra\ count rates.
The rates are well correlated and within the errors most of the
sources (14/20) are consistent with the best-fit average ratio of
rates (Swift to \chandra) of $\sim$0.19 (shown as the long dashed line in
the figure).  This is close to the ratio of effective areas at 1.5 keV
for the two observatories (135 cm$^2$ / 600 cm$^2$ $\approx$ 0.225),
which is shown as the short dashed line in the figure.  The six sources that
deviate from the average ratio of rates do so by generally no more
than about a factor of two and are likely just time variable sources.

Having demonstrated the astrometric and photometric accuracy of the
Swift data, we are now in the position to determine the X-ray count
rate and detection significance of SN~2005ke.  First we note that the
count rate of $2.2\times 10^{-4}$ s$^{-1}$ quoted by \citet{immler06}
would lead us to expect something like 25 detected photons, after
background subtraction, within our fiducial 9$^{\prime\prime}$ radius
circular region.  In fact, we detect only 15 total events, which
includes both source and background emission. We estimate the
background using the 4 regions denoted with dotted circles in 
Fig.~\ref{ke_swchha_img},
which are at least 18$^{\prime\prime}$ from the SN position, are all
equidistant ($\sim$1$^\prime$) from the center of the nuclear emission 
(as indicated by the large dashed circle) and have the same
radius as the source region.  This yields an estimate of $7.8\pm 1.4$
background events for the SN~2005ke source region. Our background
subtracted count rate (including the multiplicative PSF-correction
factor of 2) becomes $(5.7 \pm 3.3)\times 10^{-5}$ s$^{-1}$. 
Even if we vary the source extraction region (by doubling it, for 
example), the  number of source counts never exceeds the background by 
more than 2 $\sigma$. This is
not a significant detection, so we determine a count rate upper limit
of $1.6\times 10^{-4}$ s$^{-1}$ (at 3 sigma confidence) from the Swift
data.  This corresponds roughly to a \chandra\ count rate of $8\times
10^{-4}$ s$^{-1}$ (using a ratio of rates of 0.2), which is several
factors higher than the \chandra\ count rate upper limit we derived
for SN~2002bo.

\section{Limits on the Ambient CSM}

We use the model of \citet[hereafter CCL]{chug04} to determine
constraints on the density of the ambient medium surrounding SN~2002bo,
SN~2002ic, and SN~2005gj from our X-ray flux upper limits.  We do not
consider SN~2005ke, since the time of X-ray observation ($>$40 days
after explosion) and the limit on X-ray luminosity
are both less constraining of
the SN models than is the case for SN~2002bo. Here we provide a brief
summary of the calculations; the reader is referred to CCL for more
details.

The model considers interaction of the SN ejecta with spherically
symmetric, smooth circumstellar gas and treats the dynamics in a thin
shell approximation.  The evolution is followed by solving the
equations of motion numerically assuming an exponential density
profile for the SN envelope and a power law CSM density profile (which
in the case of SN~2002ic includes different indices for different
radial ranges).  Specifically, the density of SN~Ia ejecta is
described by the law $\rho=\rho_0\exp(-v/v_0)$, where $\rho_0$ and
$v_0$ are defined by the mass $M$ and kinetic energy $E$.  We adopt a
typical ejecta mass $M=1.4~M_{\odot}$, and kinetic energy
$E=1.4\times10^{51}$ ergs \citep[the model PDD3 of][]{hofkho96}.  The
outermost unburnt layers of the ejecta ($v>15000$ km s$^{-1}$) are
represented as pure oxygen.  Interior to this the ejecta are Si- and
Fe-rich, while the CS wind is assumed to be hydrogen-rich.  The
numerical model provides the time evolution of the shock radii,
velocities, and temperatures.

The internal structures of the forward and reverse shocks are not
determined in the model, so the X-ray luminosity at age $t$ is
calculated from the instantaneous kinetic luminosity of the shock,
$L_{\rm x}=\eta L_{\rm kin}$ \citep{chefra94,chu92}. The radiation
efficiency factor $\eta$ is given by $\eta=t/(t+t_{\rm c})$, where
$t_{\rm c}$ is the cooling time of the post-shock gas.  Electron and
ion temperatures are fully equilibrated for SN~2002ic and SN~2005gj
due to Coulomb collisions
(however, see below for SN~2002bo) and the shape of the emission is
given simply by a thermal bremsstrahlung spectrum at the appropriate
temperature; no line emission is included.  This is a conservative
approximation, since including line emission would boost the modeled
X-ray emission and thereby decrease our limit on the ambient gas density.

The reverse shock is essentially radiative in all cases we consider
here, so the shocked metal-rich SN ejecta create a cool dense shell
(CDS) at the contact discontinuity, which causes severe absorption of
X-ray emission from the reverse shock.  The absorption coefficient for
the metal-rich ejecta is given by $k_X \approx 5000 (E/1\, {\rm
keV})^{-8/3}$ cm$^2$ gm$^{-1}$.  X-rays are also absorbed by the CSM,
although this material is assumed to be hydrogen-rich for which $k_X$
is some 50 times smaller.  That portion of the total X-ray luminosity
absorbed by SN unshocked ejecta, CDS, and unshocked circumstellar (CS)
gas is presumed to be fully re-emitted in the optical band.

One of the most important quantities influencing the X-ray luminosity
of the SN/CSM interaction is the density of the wind.  Typically this
has been expressed in terms of a constant wind speed, $v_w$, and mass
loss rate, $\dot{M}$, as $\rho_w = \dot{M} / (4\pi v_w r^2)$.  The
quantity that we constrain with the \chandra\ observations is the wind
density parameter $w=\dot{M}/v_w$.  When quoting mass loss rates we
will give values in terms of $v_{w10} \equiv v_{w}/10\, \rm km\,
s^{-1}$.

\subsection{SN~2002bo}

It is doubtful that the shocked electrons and ions in the forward shock 
 reach temperature
equilibration in the low density CS wind around SN~2002bo. In general
the average post-shock electron temperature, $T_e$, will be in the range
$\mu_e/\bar{\mu}\leq T_e/T_i\leq 1$, where $\mu_e=m_e/m_p=1/1840$ and
$\bar{\mu}$ is the mean molecular weight.  We calculate
intermediate values of the electron temperature using the relation
$T_e / T_i = 1 - [1-(\mu_e/\bar{\mu})] [1-\exp(-\theta)]/\theta$,
where $\theta=t/t_{\rm eq}$ and $t_{\rm eq}$, the Coulomb
equilibration time, is evaluated conservatively assuming $T_e=T_i$.
This expression is the average $T_e/T_i$ for an adiabatic plane shock
in a homogeneous medium at age $t$ for a constant post-shock
equilibration timescale.  Because we use the maximal value of $t_{\rm
eq}$, the value of $T_e$ recovered in this way will be underestimated,
and accordingly, our wind density constraints will be somewhat
overestimated.

\begin{figure}
\plotone{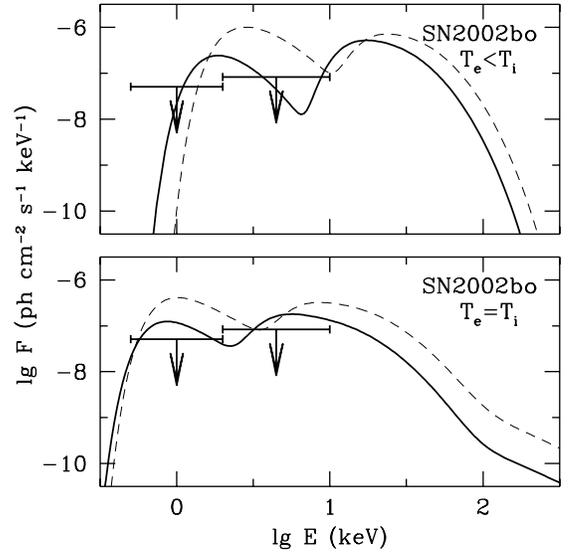}
\caption{
Model X-ray emission from SN~2002bo on day 9.3 after explosion.  The
upper panel shows the case with non-equilibrated temperatures
($T_e<T_i$) for wind density parameters of $w=1.2\times10^{15}$ g
cm$^{-1}$ (solid curve) and $w=3\times10^{15}$ g 
cm$^{-1}$ (dashed
curve), corresponding, respectively, to models  ``neq12'' and ``neq30''
in table~\ref{tab_bomod}.
The lower panel shows the model spectra with equal electron
and ion temperatures for $w=6\times10^{13}$ g cm$^{-1}$ 
(solid curve, model ``eq0.6'')
and $w=1.5\times10^{14}$ g cm$^{-1}$ (dashed curve, model ``eq1.5'').
In each panel
the solid curve corresponds to the 99\% upper limit on X-ray emission;
the dashed curve indicates how the spectra vary with increasing wind
density parameter. For comparison we plot representative \chandra\
upper limits in two broad energy bands using the absorbed
bremsstrahlung model with $N_H=2\times 10^{22}$ atoms cm$^{-2}$.  
}
\label{bo_mod}

\end{figure}

We ran a set of models with wind density parameter values in the range
$6\times10^{14}$ g cm$^{-1}$ to $3\times10^{15}$ g cm$^{-1}$ 
without assuming complete temperature equilibration
for a SN age of 9.3
d and distance of 22 Mpc \citep{kris04}. Each trial spectrum was input
to the spectral fitting program and compared to the \chandra\
data in exactly the same manner as described above (\S~\ref{s-obs}).
A value of $w=1.2\times10^{15}$ g cm$^{-1}$ was found to be the
maximum value allowed by the \chandra\ data at the 99\% confidence
level.  Fig.~\ref{bo_mod} plots this model along with a rejected case
with higher wind density parameter and, for reference, we plot the
broadband flux upper limits for the absorbed bremsstrahlung model with
$N_H=2\times 10^{22}$ atoms cm$^{-2}$.

At about 10 keV the model flux is dominated by the reverse shock
emission, while the forward shock with its cooler electrons produces
the weaker peak at about 1 keV.  The low electron temperature of the
forward shock is the result of inefficient equilibration in the low
density wind.  The electron and ion temperatures in the reverse shock,
on the contrary, are close to equilibrated.  Moreover, the reverse
shock is radiative during the initial period up to $\sim$28 days for
wind parameters in the range being considered.  The cool dense shell
of the reverse shock is responsible for the strong absorption of
X-rays from the reverse shock.

Given that the amount of collisionless electron heating at the forward
shock is highly uncertain we also ran a set of models for the
equipartition case, $T_e=T_i$. Here the maximal wind density parameter
is $w=6\times10^{13}$ g cm$^{-1}$ (Fig.~\ref{bo_mod}).  Remarkably, in
the latter case the peak at $\sim$1 keV is dominated by Comptonization
of the SN radiation by the hot electrons of the forward shock. The
Comptonized spectral component is described by a power law $\propto
E^{-\alpha}$ with $\alpha$ dependence on $\tau_{\rm T}$ and
$kT_e/m_ec^2$ taken according to \citet{PSS76}.

For reference in Table~\ref{tab_bomod} we present numerical values for
the parameters corresponding to the interaction models for SN~2002bo on
day 9.3 that are plotted in Fig.~\ref{bo_mod}.  The top two models correspond
to the non-equilibrated shocks ($T_e<T_i$), while the other two models
correspond to the equilibrated ones ($T_e=T_i$).  The columns
list: a model designation, the wind density parameter
($w$), thin shell radius ($R_{\rm s}$), velocity of the reverse
($v_{\rm RS}$) and forward ($v_{\rm FS}$) shocks, electron temperature
of the reverse ($T_{e,{\rm RS}}$) and forward ($T_{e,{\rm FS}}$)
shocks, hydrogen column density of the CS gas ($N_{\rm H}$) for
$r>R_{\rm s}$ assuming H abundance $X=0.7$. The last column gives the
mass of the cool dense shell formed by the shocked SN ejecta.  This
metal-rich matter strongly suppresses X-rays from the reverse shock and
partially absorbs X-rays from the rear side of the forward shock.

\begin{deluxetable*}{lcccccccc}
\tablewidth{0pt}
\tablecaption{SN~2002bo Model Parameters}
\tablecolumns{9}
\tablehead{
  \colhead{Model} & 
  \colhead{$w$} &
  \colhead{$R_{\rm s}$} &
  \colhead{$v_{\rm RS}$} & 
  \colhead{$v_{\rm FS}$} & 
  \colhead{$T_{e,{\rm RS}}$} & 
  \colhead{$T_{e,{\rm FS}}$} & 
  \colhead{$N_{\rm H}$} & 
  \colhead{$M_{\rm CDS}$} \\
    &
  \colhead{\small ($10^{14}$ g cm$^{-1}$)} & 
  \colhead{\small ($10^{15}$ cm)}   &  
  \colhead{\small ($10^8$ cm s$^{-1}$)}  &
  \colhead{\small ($10^8$ cm s$^{-1}$)} & 
  \colhead{\small (keV)} & 
  \colhead{\small (keV)} & 
  \colhead{\small ($10^{21}$ cm$^{-2}$)} &
  \colhead{\small ($10^{-3}~M_{\odot}$)} }
\startdata
neq12  & 12 & 2.22   &   2.68  &  24.9  & 18.3    &  1.93   &  18  &  5.6 \\
neq30  & 30 & 2.02   &   2.72  &  22.4  & 22.6    &  4.76   &  50  & 11  \\
eq0.6   & 0.6 & 2.81   &   2.18  & 32.7   &  16.5   &  1290  & 0.71 & 0.33 \\
eq1.5   & 1.5 & 2.64   &   2.41  & 30.4   &  20.1   &  1110  & 1.9  & 0.92 \\
\enddata
\label{tab_bomod}
\end{deluxetable*}

We also applied Lundqvist's models for the X-ray emission emerging
from the presumed wind surrounding a SN Ia (Lundqvist et al.~2007, in
preparation).  Note that there are some differences in assumptions and
details in this model compared to those just presented.  However our
intent here is not to resolve these differences, but rather to present
an independent estimate of the expected X-ray flux from the
interaction between SN Ia ejecta and CSM at early times.  Lundqvist's
models use a similarity solution \citep{che82a} for the evolution with
power law indices of $n=7$ in the ejecta and $s=2$ in the wind. The
velocities of the forward and reverse shock at 9.3 days after
explosion are $2.9\times 10^{4}$ km s$^{-1}$ and $7.0\times 10^{3}$ km
s$^{-1}$, respectively. The full ejecta density and temperature
structure from the similarity solution is used for calculating the
X-ray emission, assuming that the composition of the outer ejecta is
equal parts of carbon and oxygen.  We verified that the density in the
reverse shocked ejecta is sufficient to assure that the electron and
ion temperatures are equilibrated there.  In these models a radiative
cold dense shell does not form and so it is emission from the reverse
shock that dominates (and in fact these calculations do not include
forward shock emission). Only thermal emission is included; although
inverse Compton scattering could be an important contribution
\citep[see, e.g., Fig.~7 in][for the case of SN1993J]{franetal96}, we
ignore it here (which is again a conservative assumption as regards
the limit on CSM density).  Intrinsic absorption comes about only from
the overlying CSM.  Models were generated and compared to the
\chandra\ spectrum as above.  The wind density limit we obtain is
comparable to those above: $w=1.6\times10^{14}$ g cm$^{-1}$.

\subsection{SN~2002ic}

\begin{figure}
\plotone{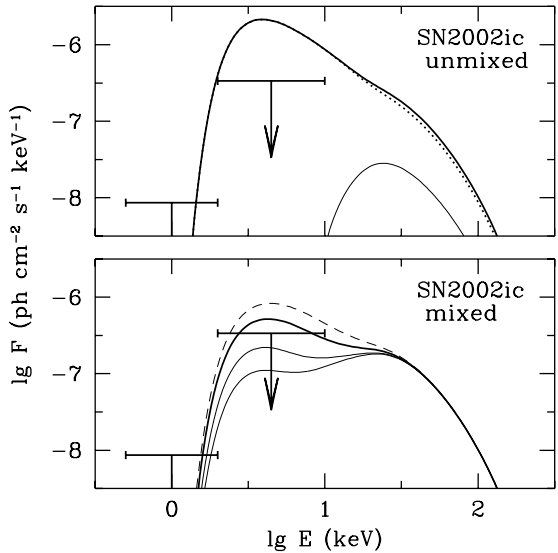}
\caption{
The predicted emergent X-ray spectrum from SN~2002ic on day 260.  For
comparison we plot representative \chandra\ upper limits in two broad
energy bands using the most absorbed bremsstrahlung model with
$N_H=8\times 10^{22}$ atoms cm$^{-2}$.  The upper panel shows the
total flux ({\em thick solid}), flux from the forward shock ({\em
dotted}), and flux from the reverse shock ({\em thin solid}) predicted
by the model of CCL (see text). This model overpredicts the
X-ray emission by a factor of $\sim$4.5. The lower panel displays the
total X-ray flux computed for the same model as above but with
fragments of cool dense ejecta homogeneously mixed in the forward
shock region for different values of the occultation optical depth:
$\tau_{\rm oc}=1$ ({\em dashed line}), $\tau_{\rm oc}=2$ ({\em thick
solid line}), $\tau_{\rm oc}=5$ ({\em thin solid line}), and
$\tau_{\rm oc}=10$ ({\em bottommost thin solid line}).  The dashed
curve is rejected by the \chandra\ observation and the $\tau_{\rm
oc} \sim 2$ case corresponds to the 99\% allowed upper limit.
}
\label{ic_mod}
\end{figure}

SN~2002ic has been studied extensively.  We take the position here that
the intense late-time ($t>50$ d) optical luminosity of SN~2002ic is
powered by the interaction of SN~Ia ejecta with a dense circumstellar
environment \citep{ham03b,wang04,chug04,nomoto05}. The CCL interaction
model that accounts for the optical luminosity and spectrum requires a
dense stellar-wind-type density profile suggesting a high mass-loss
rate ($\dot{M} \sim 10^{-2} v_{w10}\, M_\odot\, {\rm yr}^{-1}$) from
the progenitor system \citep{chuyun04}.  This model predicts an X-ray
flux of order $10^{-6}$ photon cm$^{-2}$ s$^{-1}$ kev$^{-1}$ in the
3--10 keV energy band for ages in the range $200-400$ d.  The
predicted X-ray flux on day 260 along with contributions of the
reverse and forward shocks for the standard parameter set (CCL) is
plotted in Fig.~\ref{ic_mod} for a distance 285 Mpc along with an
estimate of the \chandra\ flux upper limit (using the most absorbed
bremsstrahlung model with $N_H=8\times 10^{22}$ atoms cm$^{-2}$).  We
dub this model `unmixed' for reasons that will become clear below.
The modeled X-ray flux is clearly dominated by the emission of the
forward shock; emission from the reverse shock is absorbed by the cool
dense shell formed in the radiative ejecta.  We compared
this spectrum to the \chandra\ data and found that the X-ray flux
exceeds the observational upper limit by a factor of $\approx4.5$.
The disagreement is significant and indicates that the model is
incorrectly predicting the X-ray emission from the forward shock.

The model ignores possible deviation from spherical symmetry
suggested by polarization data \citep{wang04}. Generally, if the CS
gas is concentrated in a dense thin equatorial disk (thickness $\ll$
radius), it is conceivable that the bulk of the kinetic energy of the
interaction could be dissipated deep inside the SN envelope. As a
result, the ratio of the emergent X-ray flux to the optical flux could
be small because of strong absorption in the dense disk. However
this is unlikely to be the case.  Indeed, large deviations from global
spherical symmetry for the interaction are ruled out by the similarity
of line widths and profiles for all known SN~Ia with strong CS
interaction, i.e., 2002ic, 1999E, 1997cy (CCL). Furthermore a modest
level of asymmetry should not significantly affect the ratio of the
emergent-to-absorbed X-ray luminosity compared to the spherical case.

More serious consequences for the X-ray luminosity might result from
the omission of the internal structure of the forward shock in the thin
shell approximation.  To estimate the error introduced by this
approximation we computed a hydrodynamical model for the standard
parameter set as before (CCL) using a one-dimension hydrodynamical
Lagrangian code with artifical viscosity \citep[e.g.,][]{gull73}.  On
day 260 the radius of the contact discontinuity of the hydrodynamical
model coincides with the radius of the thin shell model to an accuracy
of 1\%.  The forward shock structure of the hydrodynamical model was
used to calculate the X-ray emission from the shocked gas assuming
instant equilibration of ion and electron temperatures.  The X-ray
luminosity of the forward and reverse shocks is lower by factors of
1.5 and 1.3, respectively, compared to the thin shell model, while the
temperature of the forward shock is greater (80 keV compared to 50 keV
in the thin shell model). The lower X-ray luminosity of both shocks in
the hydro model results in an optical luminosity lower by a factor of
1.3. To recover the optical bolometric luminosity the CS density in
the hydro model needs to be slightly larger.  We found that a
$\approx$20\% increase in the CS density is sufficient to recover the
required optical luminosity.  The new total mass of the CS envelope is
$\approx$1.9 $M_{\odot}$ within a $3\times10^{16}$ cm radius compared
to $\approx$1.6 $M_{\odot}$ found in the thin shell approximation
(CCL).  The increase of the CS density by 20\% slightly increases the
X-absorption thus reducing the peak of the X-ray flux by less than
20\%.  We thus conclude that differences between the hydrodynamical
and thin shell models are rather minor and cannot account for the
significant disparity between the \chandra\ upper limit and the
predicted X-ray flux.

Two missing factors that could potentially resolve the issue are the
Rayleigh-Taylor (RT) instability of the CDS and clumping of the CS
gas. The RT instability of the shocked decelerated ejecta is a generic
attribute of the SN/CS interaction \citep{che82b}, while the clumpiness
of the CS matter around SN~2002ic has been invoked already to account
for the H$\alpha$ line profile (CCL).  The RT instability leads to
fragmentation and mixing of CDS fragments with the hot gas of the
forward shock 
\citep{cheblo95,bloell01}.  Clumpiness of the CS matter favors more efficient
mixing of the CDS fragments within the forward shock principally
through two mechanisms.  First, the CS clouds penetrate deep inside
the intercloud shock before they get shocked, fragmented and mixed
\citep{KMC94,blo01}. As a result the bulk of the kinetic energy
related to CS clouds dissipates closer to the contact discontinuity.
Second, the CS clouds engulfed by the intercloud shock generate vortex
turbulence that favors more efficient penetration of the RT spikes in
the forward shock \citep{JJN96}.  We thus propose a scenario for
SN~2002ic in which the forward shock is no longer a regular layer of
hot gas but instead is a macroscopic {\em mixture of hot gas and cool
metal-rich fragments} of the CDS.  The major outcome of this
modification should be an additional component of absorption of the
X-ray emission from the forward shock by the intermixed fragments of
metal-rich CDS.

To illustrate this effect we will calculate the absorption of X-rays
from the forward and reverse shocks assuming the same thin shell model
as in CCL, but with the following corrections.  The CDS is assumed to
be fragmented and homogeneously mixed in the forward shock
layer. Remarkably, the analysis of the optical spectrum of SN~2002ic
suggests that the CDS is indeed fragmented and well mixed. The ratio
of the cumulative area of clumps to the area of the spherical shell,
which is a measure of mixing, was estimated to be $S/S_0\sim50$ on day
234 (CCL).  The occultation optical depth of the fragmented CDS, i.e.,
the average number of fragments on the line of sight in the shell for
randomly oriented plane fragments is then $\tau_{\rm
oc}=0.5(S/S_0)\sim25$.  Small values of the occultation optical depth,  
$\tau_{\rm oc} \sim 1-2$, would mimic the possible presence of ``holes''
in the mixing zone due to strong angular variation in the column density 
\citep{cheblo95}.
Another relevant value is the average optical
depth for X-rays $\tau$, which is determined by the average column
density of CDS material and the X-ray absorption coefficient, for
which we take the metal-rich case as given above. The effective
optical depth of the ensemble of fragments is then $\tau_{\rm
eff}=\tau_{\rm oc}[1-\exp(\tau/\tau_{\rm oc})]$.  In the limit
$\tau_{\rm eff}\gg 1$ the effect of mixing for a homogeneous thin
spherical layer with unabsorbed luminosity $L_0$ is to reduce the
emergent X-ray luminosity to a value of $L_0/(4\tau_{\rm
eff})$. Indeed, only half of the photons emitted in a layer with
thickness equal to the mean free path ($\Delta R/\tau_{\rm eff}$) can
escape.  The other factor of one-half comes from averaging over the
angle between the photon direction and surface normal.

The mixing we invoke may, however, be incomplete, which can be
mimicked by adopting a lower occultation optical depth of the
homogeneous layer, $\tau_{\rm oc}<25$. We also take into account
additional absorption by the fragmented shocked CS clouds with normal
composition, assuming that 50\% of the swept-up CS gas resides in the
cool shocked cloud material with $\tau_{\rm oc, cs}=1$. This
component, presumably responsible for the H$\alpha$ emission, produces
negligible X-ray absorption.

The computed total flux of escaping X-rays is presented in the lower panel of
Fig.~\ref{ic_mod} for $\tau_{\rm oc}=1$, 2, 5 and 10, which vary from
low to high degrees of mixing of CDS fragments into the forward shock
region. The plot shows a pronounced effect on the X-ray absorption
(see Fig.~\ref{ic_mod}) by the macroscopically mixed CDS
material. The case $\tau_{\rm oc}=2$ corresponds to the maximal flux
tolerated by the \chandra\ upper limit.  This study indicates that
mixing of the shocked SN ejecta with the hot gas of the forward shock
can resolve the X-ray flux disparity between the original model that
explains the optical spectrum (CCL) and the new \chandra\ observations
of SN~2002ic.

\subsection{SN~2005gj}

\begin{figure}
\plotone{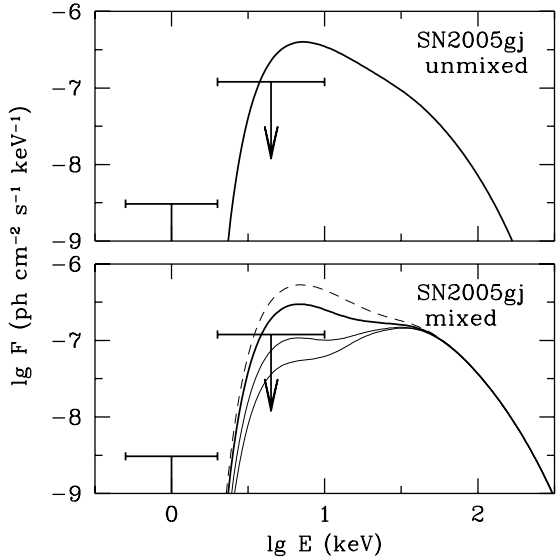}
\caption{
The model X-ray spectrum of SN~2005gj.  For comparison we plot
representative \chandra\ upper limits in two broad energy bands using
the highly absorbed bremsstrahlung spectrum with $N_H=8\times 10^{22}$
atoms cm$^{-2}$.  The upper panel shows the unmixed model used
previously for SN~2002ic and recomputed at an epoch of $t\sim80$ d as
appropriate for SN~2005gj. This model overpredicts the X-ray emission
by a factor of $\sim$4. The lower panel displays the total X-ray flux
computed for the same model as above but with fragments of cool dense
ejecta homogeneously mixed in the forward shock region for different
values of the occultation optical depth: $\tau_{\rm oc}=1$ ({\em
dashed line}), $\tau_{\rm oc}=2$ ({\em thick solid line}), $\tau_{\rm
oc}=5$ ({\em thin solid line}), and $\tau_{\rm oc}=10$ ({\em
bottommost thin solid line}).  The dashed curve is rejected by the
\chandra\ observation and the $\tau_{\rm oc} \sim 2$ case corresponds to
the 99\% allowed upper limit.
}
\label{gj_mod}
\end{figure}

According to its early spectrum and photometry, SN~2005gj is remarkably
similar to SN~2002ic and very likely is a member of this new class of
bright Ia SNe embedded in dense CS envelope.  This position has been
strengthened by the recent light curve and spectral study of
\citet{aldetal06}. Their V- and I-band light curves for SN~2005gj match
the shape of the early time evolutionary models of \citet{chuyun04}
for SN~2002ic, although the models need to be scaled upward in
brightness by 0.5 mag or so.  The CSM interaction in SN~2005gj is
evidentally stronger than in SN~2002ic.  In the absence of a detailed
hydrodynamical model for SN~2005gj, we utilize the same model and
parameter values as for SN~2002ic.
First we computed the X-ray flux at the \chandra\ observation epoch
($t\sim 80$ d) neglecting the effects of ejecta mixed into the forward
shock.  The resulting spectrum is presented in the upper panel of
Fig.~\ref{gj_mod} for a distance to SN~2005gj of 266 Mpc together with
an estimate of the \chandra\ broad band flux upper limits (using the
most highly absorbed bremsstrahlung spectrum with $N_H=8\times
10^{22}$ atoms cm$^{-2}$).  The model spectrum overpredicts the X-ray
flux by a factor of 4. Although little is currently published about
the environment of SN~2005gj, the \citet{aldetal06} study suggests that
the density of the ambient medium is, if anything, higher than that
around SN~2002ic, which exacerbates the discrepancy with the \chandra\
flux limit.  Therefore we next considered the mixing of CDS fragments
into the forward shock region in the case of SN~2005gj. The required
occultation optical depth to not exceed the \chandra\ limit is
$\tau_{\rm oc}<2$ (Fig.~\ref{gj_mod}).  Consequently, if the CS
environment around SN~2005gj is as dense as that around SN~2002ic,
mixing of the fragmented CDS must be already significant at this early
epoch.

\section{Discussion and Conclusion}

With our \chandra\ observation of SN~2002bo we have set the most
sensitive X-ray flux upper limits at an earlier epoch than for any
previous SN~Ia. The previous best case was that of SN~1992A for which a
\rosat\ upper limit of $\sim$$10^{-14}$ ergs cm$^{-2}$ s$^{-1}$ was
set on the 0.5-2 keV band X-ray flux $\sim$35 days after explosion
\citep{schpet93}.  Our \chandra\ limit in the same energy band is an
order of magnitude lower and was set only 9 days after explosion.  We
also set a sensitive upper limit in the important 2--10 keV band.

Converting our flux upper limits to constraints on the density of the
CSM in the system depends on uncertain assumptions about the
thermodynamic state of the hot plasma in the expanding ejecta and
shocked wind.  As we have shown in this paper there is roughly an
order of magnitude difference in the inferred wind density under the
assumption of fully equilibrated electron and ion temperatures vs.\
the case with non-equilibrated temperatures.  The latter case yields
the more conservative (i.e., higher) constraint: a wind density
parameter $w < 1.2 \times 10^{15}$ g cm$^{-1}$.  In terms of a slow
wind with velocity of $v_{w10}=10$ km s$^{-1}$ this corresponds to a
upper limit on the mass-loss rate of $\dot{M} < 2 \times
10^{-5}~M_{\odot}$ yr$^{-1}$. The much higher sensitivity of the
\chandra\ data notwithstanding, this value is {\em larger} by a factor
of 10 or so than the mass loss upper limit derived from \rosat\ data
by \citet{schpet93} for SN~1992A.  This discrepancy is due to the
simplicity of the model used by \citet{schpet93} to evaluate their
upper limit: they do not include absorption by residual wind material
above the forward shock, assume electron-ion temperature
equilibration, compare to model luminosities without applying a
bolometric correction, and calculate the SN age from the time of
maximum rather than from the time of explosion.  In other comparisons
our \chandra\ upper limit is comparable to those found previously from
limits on the H$\alpha$ flux for the normal type Ia supernovae
SN~1994D \citep{cumm96} and SN~2001el \citep{matetal05}.  In general
limits on $\dot{M}$ of $\sim$$10^{-5}~M_{\odot}$ yr$^{-1}$ are not
stringent enough to rule out the class of symbiotic-type binaries as
SN~Ia progenitors at least not for these particular cases.

The value of these results lies in our ability to calculate, using
well understood physics, the expected X-ray emission from hot gas.
Viewed in this light, we briefly discuss limits on the CSM obtained
from radio nondetections of a number of nearby SNe~Ia
\citep{panagetal06}.  The radio results rely on semi-empirical
parameterized functional forms for the time- and frequency-dependence
of synchrotron emission and free-free absorption, whose relevant
parameters are assumed to have values given by radio results for
SNe~Ib/c. Likewise the essential parameter, i.e., the one linking the
wind density parameter to the radio luminosity of the SN, cannot be
calculated from theory with any accuracy, and an empirical
calibration, again drawn from measurements of SNe~Ib/c, must be used.
Although the sensitive radio flux limits clearly argue for low density
environments around SN~Ia, the extremely low numerical limits on the
mass-loss rates claimed by \citet{panagetal06} cannot yet be
considered definitive.

The other normal SN~Ia we study here is SN~2002ke.  We re-examine the
claim by \citet{immler06} of a tentative X-ray detection by Swift and find 
that we cannot substantiate it. We pay particular attention to the
astrometric and photometric calibration of the Swift X-ray data by
comparing to a \chandra\ observation done several months later.  We
find no evidence for a significant X-ray detection of SN~2005ke by
either Swift or \chandra, to a flux limit that is several factors
higher than what we obtained for SN~2002bo.  Since this limit is at a
much later epoch, when the intensity of the CSM/ejecta interaction
should be much reduced, we did not attempt to determine numerical
limits to the wind density parameter for this SN.

We have also presented \chandra\ upper limits for the two 
examples of SNe Ia with clear evidence for circumstellar interaction:
SN~2002ic and SN~2005gj.  The upper limit on the X-ray luminosity of
SN~2002ic in the $0.5-6$ keV band unexpectedly reveals a serious
drawback to the interaction model proposed previously (CCL): the
predicted X-ray flux turns out to be larger by at least a factor of
four.  We identified the major missing element of the model responsible
for the controversy: macroscopic mixing of cool metal-rich ejecta
fragments into the hot gas of the forward shock, which results in 
strong absorption of the X-rays emitted by the forward shock.
Interestingly, the absorption of X-rays by mixed shocked ejecta should
have the effect of decreasing the required CS density in the model to
explain the late time optical luminosity. This effect together
with the higher radiation efficiency of the interaction with a clumpy
CS matter should result in a slightly higher expansion velocity of the
shocked SN ejecta in the interaction model. These outcomes appear to
be preferred by the optical spectra of SN~2002ic (see CCL).

SN~2005gj appears to belong to the SN~2002ic-like family of SN~Ia with
dense CS envelopes \citep{aldetal06,prietoetal07}. The \chandra\ upper limit from
this object at about day 80 is a factor of four larger than the
flux predicted by the interaction model of SN~2002ic recomputed for
the corresponding epoch.  Although the strength of the interaction
argues against it, one possible explanation is that the CS
density around SN~2005gj is just somewhat lower than around SN~2002ic.
A better possibility invokes mixing of the shocked ejecta, which can
reduce the emergent X-ray flux. Which of these is more likely to be
the correct explanation requires a better understanding of the
environment of SN~2005gj than we have at present. Future studies of
the optical light curve and spectra of this interesting object are
strongy encouraged.

Resolving the issue of the X-ray non-detection of SN~2002ic-like
objects has its dark side due to our introduction of the parameter
$\tau_{\rm oc}$, which is essentially incalculable.  Even
three-dimensional hydrodynamical simulations are unlikely to be able
to recover this value in a fully self-consistent way. We, therefore,
can predict in detail neither the spectrum nor the evolution of the
X-ray flux from SN~2002ic-like supernovae. Two relevant remarks can be
made, however. First, in the case of SN~2002ic the model without
mixing predicts an increasing X-ray flux in the band below 10 keV up
until day $\sim$400 (CCL). Therefore, X-ray detection at earlier
epochs is unlikely to be more favorable, as the lack of detection of
SN~2005gj at roughly 80 days after explosion (versus 260 days for
SN~2002ic) tends to support.  The second remark relates to the
spectrum of the emergent X-ray emission: hard X-rays (i.e., with
photon energies greater than $\sim$20 keV) are not affected by the
absorption that mutes the lower energy flux (see Fig.~\ref{ic_mod}).
Future sensitive X-ray observations covering a wider energy band
($1-30$ keV), therefore, should reveal a SN~2002ic-like event as a
strongly absorbed hard X-ray source and thus verify the proposed
mixing scenario.

\acknowledgments

We appreciate the amount of effort required to successfully execute
target of opportunity observations with \chandra\ and we gratefully
acknowledge the entire \chandra\ operations staff for their rapid
response to our observation requests.  We thank the CXC Director,
Harvey Tananbaum, for awarding discretionary time to observe SN~2002ic
and SN~2005gj.  We acknowledge Carles Badenes, Andrew Baker, and
Daniela Calzetti for helpful discussions on various aspects of this
project. We also thank Stefan Immler for discussions about the Swift
observations of SN~2005ke.
This research has made use of the NASA/IPAC Extragalactic
Database (NED) which is operated by the Jet Propulsion Laboratory,
California Institute of Technology, under contract with the National
Aeronautics and Space Administration.  We also made use of data
obtained from the High Energy Astrophysics Science Archive Research
Center (HEASARC), provided by NASA's Goddard Space Flight Center.
Financial support was provided by the National Aeronautics and Space
Administration through \chandra\ Award Number GO2-3068X issued to
Rutgers University by the \cxo\ Center, which is operated by the
Smithsonian Astrophysical Observatory for and on behalf of the
National Aeronautics Space Administration under contract NAS8-03060.

{\it Facilities:} \facility{CXO (ACIS-S)}, \facility{Swift (XRT)}

\appendix

\section{X-ray Emission from Host Galaxies}

Only the host galaxies of the two nearby SNe, NGC 3170 and NCG 1371,
show any evidence for X-ray emission.  In each case there is a
compact, spectrally hard, nuclear X-ray source in addition to faint
diffuse emission. Here we present a brief report on the new
information from the \chandra\ observations of these galaxies.

\subsection{NGC~3190 and other nearby galaxies}

NGC~3190 is a member of the Hickson Compact Group 44, also the Leo
III group, and is classified as a Low Ionization Nuclear Emission
Region (LINER) galaxy. X-ray emission from this galaxy was first
reported by \citet{piletal95} using a short (5 ks) \rosat\ PSPC
observation from which it was only possible to determine a broadband
luminosity of $4\times 10^{39}$ ergs s$^{-1}$ (0.07--3 keV band), where
we have corrected their published result for our assumed
distance to the  galaxy.
\citet{liubre05} report the presence of a compact nuclear
X-ray source with an estimated X-ray flux of $8.8\times 10^{39}$ ergs
s$^{-1}$ (0.3--8 keV band) based on \rosat\ HRI
data.

The \chandra\ data (Fig.~\ref{bo_soft_img}) reveal that the compact
nuclear source is spectrally hard and, in addition, is embedded in a
faint, diffuse, spectrally soft, extended component. This latter
component tends to lie north and east of the nucleus, and clearly
avoids the prominent dust lane that runs through the galaxy from
southeast to northwest some 10$^{\prime\prime}$ south of the nucleus.
The spectra shown in Fig.~\ref{bo_spec} were extracted from the 5
regions shown on Fig.~\ref{bo_soft_img}.  We have already discussed
the bottommost spectrum in this figure (i.e., the background spectrum)
in \S 3.   
All fits were carried out with the likelihood figure-of-merit function
used for the background model fits, which does not yield an explicit
goodness-of-fit criterion.  However, the models, at least visually,
provide a fairly good description of the spectral data.
All spectral
fits, except the absorbed nuclear power-law component, include
absorption fixed at the Galactic HI column density value of $N_{\rm H}
= 2.1\times 10^{20}\, \rm cm^{-2}$.

The top spectrum comes from the innermost region: a circle
1.5$^{\prime\prime}$ (160 pc) in radius centered on the mean position
of the hard nuclear source (R.A.=10:18:05.63, Decl.=+21:49:56).  For
an acceptable fit the spectrum requires two components: a highly
absorbed power-law as well as thermal emission (using the ``mekal''
model in xspec). The best fit thermal emission model has $kT = 0.76\pm
0.05$ keV and emission measure, $n_e n_H V = 4.3\times 10^{61}$
cm$^{-3}$. The power-law component has a fixed photon index of $\Gamma
= 1.4$ with a best-fit intrinsic absorption of $N_{\rm H} = 1.7\pm0.3
\times 10^{23}$ cm$^{-2}$ and flux density at 1 keV of $F_E = 6.2\pm
1.4 \times 10^{-5}$ photons keV$^{-1}$ cm$^{-2}$ s$^{-1}$.  The
unabsorbed flux (0.5--10 keV) of the power-law component is $5.5
\times 10^{-13}$ ergs cm$^{-2}$ s$^{-1}$ or a luminosity of $3 \times
10^{40}$ ergs s$^{-1}$.  It is clear that the LINER nature of this
galaxy is due to an obscured, low luminosity, active galactic nucleus (AGN).

The three spectra in the middle of Fig.~\ref{bo_spec} are of the
spectrally soft, extended emission and all are well described by
thermal emission alone. The red spectrum was extracted from the second
smallest region on Fig.~\ref{bo_soft_img}, an ellipse with semi-major
axis lengths of 3.9$^{\prime\prime}$ and 3.2$^{\prime\prime}$ (420 pc
$\times$ 340 pc).  The best fit values for the thermal emission were
$kT = 0.80\pm 0.05$ keV and $n_e n_H V = 3.5\times 10^{61}$ cm$^{-3}$.
The green spectrum came from the next larger region: an ellipse with
semi-major axis lengths of 9.6$^{\prime\prime}$ and
8.4$^{\prime\prime}$ (1.0 kpc $\times$ 0.9 kpc).  This was well fit by
thermal emission alone with a best fit $kT = 0.74\pm 0.08$ keV and
$n_e n_H V = 2.9\times 10^{61}$ cm$^{-3}$.  Finally the blue spectrum
came from the next region: an ellipse with semi-major axis lengths of
45$^{\prime\prime}$ and 25$^{\prime\prime}$ (4.8 kpc $\times$ 2.7
kpc). Three obvious point sources were excluded from this region.
This was well fit by thermal emission alone with a best fit $kT =
0.37\pm 0.06$ keV and emission measure, $n_e n_H V = 4.2\times
10^{61}$ cm$^{-3}$.

From the spectral extraction regions we estimate emitting volumes
(assuming an ellipsoidal geometry with the line-of-sight depth equal
to the mean of the axis lengths of the extraction regions) and then
convert the emission measures given above into the density of the
emitting plasma (for $n_e / n_H = 1.2$). We find density values that
vary from $n_H = 0.27$ cm$^{-3}$ near the nucleus through values of
0.069 cm$^{-3}$ and 0.015 cm$^{-3}$ for the inner and mid galaxy
regions to a value of 0.0025 cm$^{-3}$ for the outer galaxy. These
values are broadly consistent with an r$^{-2}$ profile perhaps
pointing toward an outflow from the central regions of the galaxy.  In
summary, the X-ray emission properties of NGC~3190 closely resemble
those of other LINER galaxies studied by \citet{teretal02}: a low
luminosity and obscured AGN with soft ($kT\sim 0.8$ keV) extended
thermal emission.

X-ray emission from two other galaxies (NGC~3193 and NCG~3185) in HCG
44 were also reported by \citet{piletal95}.  We also detect X-ray emission
from these same galaxies in the \chandra\ data, which, thanks to a
fortuitous value of the roll angle, happen to lie within the field of
view. Although clearly detected, the emission from NGC 3193 falls
near the gap between chips S3 and S4, which renders derived results
somewhat inaccurate.  We do not consider this galaxy further.

NCG 3185 is $\sim$9$^\prime$ off-axis where the imaging quality of
\chandra\ is modest ($\sim$5$^{\prime\prime}$).  The PSF there is
however sufficient to separate the compact nuclear emission from the
fainter off-nuclear source, XMMU J101737.4+214144, discovered by
\citet{fosetal02}.  The published \xmm\ luminosity for the off-nuclear
source is $1.3 \times 10^{39}$ ergs s$^{-1}$ (0.5--10 keV band)
assuming a power-law source with a photon index of 2.0 and distance of
21.3 Mpc.  We detect this source in the \chandra\ data with a net
total of $21 \pm 5$ counts above background from which we estimate a
luminosity of $7 \times 10^{38}$ ergs s$^{-1}$ (in the same spectral
energy band using the same spectral form as the \xmm\ study). This
suggests some variability in the X-ray luminosity of this off-nuclear
source. The nuclear source is stronger (detected at $72 \pm 9$ counts
above background) and generates a luminosity of $2 \times 10^{39}$ ergs
s$^{-1}$ (again for the same band and spectral form).  This agrees
with the $L_X$ value published by \citet{cappietal06} from \xmm\
observations.

\subsection{NGC~1371}

NGC1371 has typically been classified \citep[see, e.g.,][]{esk02} as a
weak bar, early-type spiral galaxy (SAB(rs)a according to the RC3).
It is known to display an ``Extended Nuclear Emission-line Region''
(ENER) in the form of a disk-shaped zone of H$\alpha$ emission
\citep{hamdev99} extending over an elliptical region with major axis lengths
of roughly
30$^{\prime\prime}$ $\times$ 13$^{\prime\prime}$ (2.5 kpc $\times$ 1.1
kpc) (see Fig.~\ref{ke_swchha_img}).  A compact nuclear source
appears in H$\alpha$ as well.  The total H$\alpha$ luminosity of the
galaxy is $L_{{\rm H}\alpha} = 4.9 \times 10^{40}$ ergs s$^{-1}$,
which is uncorrected for extinction and therefore a luminosity lower
limit.  The nuclear component (i.e., the emission within 1 kpc
$\approx$ 12$^{\prime\prime}$, which contains much of the ENER
emission) contains roughly 6\% of the total H$\alpha$ luminosity. The
nucleus of this galaxy has not been spectroscopically
identified. 

Our \chandra\ data reveal a compact source of X-ray emission at
R.A.=03:35:01.35, Decl.=$-$24:55:59.6 that is coincident with the
nucleus and is well described by an absorbed power law spectrum (see
top curve and data points in Fig.~\ref{ke_spec}).  The best-fit
parameter values are a photon index of $\Gamma = 1.5^{+0.3}_{-0.2}$,
intrinsic absorption of $N_{\rm H} = 2.5^{+0.6}_{-0.4} \times 10^{22}$
cm$^{-2}$ (considerably higher than the Galactic column toward
NGC~1371), and flux density at 1 keV of $F_E = 1.6^{+1.0}_{-0.5}
\times 10^{-4}$ photons keV$^{-1}$ cm$^{-2}$ s$^{-1}$.  The unabsorbed
flux (0.5--10 keV) is $1.2 \times 10^{-12}$ ergs cm$^{-2}$ s$^{-1}$
corresponding to a luminosity of $4 \times 10^{40}$ ergs
s$^{-1}$. Whether the excess absorption is intrinsic to the nuclear
source or comes from the intervening foreground portion of the galaxy
is not clear.  The nuclear X-ray and H$\alpha$ luminosity values fall
right on the published correlation between these quantities for
Seyfert galaxies, LINERs, and AGN from \citet{hoetal01}.  Thus we
conclude that NGC~1371 contains a low luminosity AGN with properties
similar to those found in other nearby spiral galaxies.

We also detect diffuse X-ray emission on larger scales.  The red
spectrum in Fig.~\ref{ke_spec} (second from top) comes from the inner
elliptical region identified on Fig.~\ref{ke_soft_img}, which
corresponds well to the extended H$\alpha$ disk. The spectral
extraction region used here has major axis lengths of
31.4$^{\prime\prime}$ $\times$ 16.1$^{\prime\prime}$ (2.6 kpc $\times$
1.3 kpc) in size.  Fits to the X-ray spectrum from this region prefer
a two component model: low temperature thermal emission ($kT \approx
0.3$ keV) in addition to either a hot thermal ($kT \approx 10$ keV) or
power law ($\Gamma \approx 1.8$) component.  Note that for these fits
the absorbing column was fixed to the Galactic value (i.e., no extra
absorption intrinsic to NGC 1371 was required).  The unabsorbed
0.5--10 keV band flux of the hard component alone (for either spectral
form) is $4.0 \times 10^{-14}$ ergs cm$^{-2}$ s$^{-1}$ corresponding
to a luminosity of $1.4 \times 10^{39}$ ergs s$^{-1}$.  The soft
component yields a flux of $1.6 \times 10^{-14}$ ergs cm$^{-2}$
s$^{-1}$ (equivalent to a luminosity of $5.6 \times 10^{38}$ ergs
s$^{-1}$) entirely in the 0.5--2 keV band.

Faint diffuse soft X-ray emission is visible on Fig.~\ref{ke_soft_img}
within the outer elliptical region centered on the nucleus.  This
region, with major axis lengths of 104$^{\prime\prime}$ $\times$
75$^{\prime\prime}$ (8.6 kpc $\times$ 6.2 kpc), covers the bright
central portion of NGC~1371. Fits to the X-ray spectrum (shown as the
blue curve and data points third from the top on Fig.~\ref{ke_spec})
are consistent with a thermal spectrum ($kT \approx 0.4$ keV). The
X-ray flux is $2.8 \times 10^{-14}$ ergs cm$^{-2}$ s$^{-1}$
(equivalent to a luminosity of $1.0 \times 10^{39}$ ergs s$^{-1}$) and
again appears entirely in the 0.5--2 keV band.

The fitted emission measure of the soft thermal component in the disk
region is $n_e n_H V \sim 3\times 10^{61}$ cm$^{-3}$, while that of
the outer elliptical region is $4\times 10^{61}$ cm$^{-3}$.  These
values lead to mean gas densities of $n_H = 0.015$ cm$^{-3}$ in the
disk region and $n_H = 0.0024$ cm$^{-3}$ within the outer ellipse.
For the inner disk-like region of NGC~1371, the close morphological
match between the X-ray and H$\alpha$ emission argues that the X-rays
we see surrounding the nuclear AGN come from the integrated flux of
sources linked to recent star formation, such as supernova remnants,
winds from young massive stars, X-ray binaries, and so on.  The
composite nature of the X-ray spectrum, especially the presence of the
hard emission component, points to the need for a contribution from a
population of X-ray binaries.  However no individual X-ray source is
clearly visible in this region (aside from the nucleus) with a
luminosity greater than $\sim$$10^{38}$ ergs s$^{-1}$.

There are three unresolved X-ray sources that are nearly coincident
(i.e., offset by $<$3$^{\prime\prime}$) with individual H II regions
in the H$\alpha$ image of \citet{hamdev99}.  The brightest of these,
CXOU~J033506.0$-$245713, is approximately 1.6$^\prime$ southeast of
the nucleus and is indicated by the cross in Fig.~\ref{ke_swchha_img}
(right panel).  The 0.5--10 keV band flux of this source is $3 \times
10^{-14}$ ergs cm$^{-2}$ s$^{-1}$ assuming a hard power-law or high
temperature bremsstrahlung spectral form, either of which describes
the data well.  The other two sources, CXOU~J033500.2$-$245437 and
CXOU~J033452.7$-$245527, are each roughly one-third as bright in the
\chandra\ data.  The luminosities of these sources, if they are in
NCG~1371 as seems likely, are $3\times 10^{38}$ ergs s$^{-1}$ for the
two fainter ones and $10^{39}$ ergs s$^{-1}$ for the brighter one.
With the exception of the nucleus, these are the brightest sources
within the optical extent ($5.6^\prime \times 3.9^\prime$) of NGC~1371.
Their high X-ray luminosity, generally hard spectra, and spatial
association with H II regions argue that these are accreting high mass
X-ray binaries.

Four other sources were detected at high significance ($\ge$10 counts)
within this area, while there were three others with 7--8 counts.
None of these other sources shows a counterpart in the H$\alpha$ or
continuum images of \citep{hamdev99} or in 2MASS infrared images
\citep{jaretal03}.  We expect only $\sim$3 unrelated background
sources at these flux levels, so most of these should be new sources
associated with NCG~1371.

\end{document}